\documentclass[rmp,superscriptaddress,twocolumn,10pt]{revtex4-2}

\usepackage{titlesec}
\usepackage{textgreek}
\usepackage{amssymb}
\usepackage{amsmath}
\usepackage{graphicx}
\usepackage{color}
\usepackage{xcolor}
\usepackage[normalem]{ulem}
\usepackage{float}
\usepackage{setspace}
\usepackage{multirow}
\usepackage{tabularx}
\usepackage{enumerate}
\usepackage[bb=boondox]{mathalpha}
\usepackage{bm}
\usepackage[a4paper]{geometry}

\setcitestyle{super}

\newcommand*\xbar[1]{%
  \hbox{%
    \vbox{%
      \hrule height 0.3pt 
      \kern0.3ex
      \hbox{%
        \kern-0.1em
        \ensuremath{#1}%
        \kern-0.1em
      }%
    }%
  }%
}

\titlespacing*{\section}{0.ex}{4ex}{0.ex}

\makeatletter
\def\hlinewd#1{%
\noalign{\ifnum0=`}\fi\hrule \@height #1 %
\futurelet\reserved@a\@xhline}
\makeatother

\makeatletter
\def\frontmatter@above@affiliation@script{\addvspace{8.0\p@}}
\makeatother

\newcommand{\rev}[1]{{\color{blue} #1}}

\begin{document}

\begin{spacing}{1.}

\title{Network clique cover approximation to analyze complex contagions through group interactions}

\author{\large Giulio Burgio}
        \affiliation{Departament d'Enginyeria Inform\`{a}tica i Matem\`{a}tiques, Universitat Rovira i Virgili, Tarragona, Catalonia, Spain}
\author{\large Alex Arenas}
        \affiliation{Departament d'Enginyeria Inform\`{a}tica i Matem\`{a}tiques, Universitat Rovira i Virgili, Tarragona, Catalonia, Spain}
\author{\large Sergio G\'{o}mez}
        \email{sergio.gomez@urv.cat}
        \affiliation{Departament d'Enginyeria Inform\`{a}tica i Matem\`{a}tiques, Universitat Rovira i Virgili, Tarragona, Catalonia, Spain}
\author{\large Joan T.\ Matamalas}
        \affiliation{Center for Interdisciplinary Cardiovascular Sciences, Cardiovascular Division, Department of Medicine, Brigham and Women’s Hospital, Harvard Medical School, Boston, MA 02115, USA}

\maketitle

\section*{\large{Abstract}}

{\fontfamily{cmss}\selectfont Contagion processes have been proven to fundamentally depend on the structural properties of the interaction networks conveying them. Many real networked systems are characterized by clustered substructures representing either collections of all-to-all pair-wise interactions (cliques) and/or group interactions, involving many of their members at once. In this work, focusing on interaction structures represented as simplicial complexes, we present a discrete-time microscopic model of complex contagion for a susceptible-infected-susceptible dynamics. Introducing a particular edge clique cover and a heuristic to find it, the model accounts for the higher-order dynamical correlations among the members of the substructures (cliques/simplices). The analytical computation of the critical point reveals that higher-order correlations are responsible for its dependence on the higher-order couplings. While such dependence eludes any mean-field model, the possibility of a bi-stable region is extended to structured populations.
}

\section*{\large{Introduction}}
\label{sec:intro}

Epidemics \cite{anderson1992infectious}, rumor spreading \cite{daley1965stochastic}, adoption \cite{rogers2010diffusion} and opinion dynamics \cite{katz1966personal} are well-known manifestations of real-world contagion processes. One of the most remarkable achievements of network science has been the characterization of the dependence of contagion processes on the structural properties of the interaction network through which they spread \cite{valente1995network,cowan2004network,pastor2001epidemic,watts2007influentials}.  In particular, many real-world networks of interest, especially the social ones, boast a clustered and cycles-rich structure. Triadic closures are indeed renowned to be a distinguishing feature of social systems \cite{watts1998collective,bianconi2014triadic}, together with the presence of larger communities in which every element is connected to (nearly) any other element in them \cite{palla2005uncovering}. Households and workplaces are common contact-based examples of that, while online social communities and groups in messaging apps are information-based ones. In the language of graph theory, such all-to-all substructures are called cliques. Specifically, a $n$-clique (or clique of order $n$) consists of $n$ nodes all pair-wisely connected to each other, i.e., a complete subgraph of $n$ nodes.

Apart from a collection of dyadic interactions, cliques can be also regarded as the pair-wise projection of richer substructures representing group interactions (also known as `higher-order' interactions), involving more than two agents (nodes) at once. In fact, from few years on, a growing branch of literature dedicated to the study of various dynamical processes involving group interactions \cite{battiston2020networks,burgio2020evolution,dai2020d,ghorbanchian2020higher,tadic2021hidden,sun2020renormalization,andjelkovic2020topology}, has been showing that such interactions can heavily affect the dynamics, and neglecting them can therefore lead to wrong predictions.

The interaction patterns can be properly formalized by means of hypergraphs \cite{bretto2013hypergraph, lambiotte2019networks}, a generalization of graphs in which the nodes can be grouped in hyperedges of any order ---not only in pairs. Specifically, here we make use of simplicial complexes \cite{jonsson2007simplicial}, defined as sets of faces, whereby a face is a set of nodes with the hereditary property, stating that each of its subsets is also a face of the simplicial complex (SC). Associating a group interaction to each face, a SC then represents the entire set of interactions among the nodes in it.

Going to the dynamics, we are interested in complex contagion processes \cite{granovetter1978threshold,centola2007complex,centola2010spread,melnik2013multi,watts2002simple}, in which the outcome of a potentially spreading interaction depends on how many different contagious agents take part to it, and not ---as in simple contagions--- only on the strength of the interaction. The presence of complex contagion mechanisms has been assessed in various contexts \cite{guilbeault2018complex}, but largely in online social networks and forums \cite{ugander2012structural,weng2012competition}, thanks to their unique data traceability.

Several studies have recently appeared showing the dynamical effects of group interactions on complex contagion. They provide either a qualitative understanding by means of mean-field approximations \cite{iacopini2019simplicial,landry2020effect,st2021bursty}, finding such interactions as responsible for critical mass effects; or a more quantitative one, reinforcing the previous qualitative findings, but for very particular hypergraphs \cite{bodo2016sis,jhun2019simplicial,de2020social}. All the studies considered continuous-time dynamics and, noteworthy, uncorrelated nodes' states. For the case of $2$-dimensional SCs (i.e., consisting of edges and triangles), Matamalas et al.\ \cite{matamalas2020abrupt} provided a microscopic discrete-time model accounting for first-order (two-nodes) and, only partially, second-order (three-nodes) dynamical correlations. However, as shown afterwards, the same subtle inconsistencies that make the model applicable to any $2$-dimensional SC, also impede the computation of the critical point.

In this work, looking for a discrete-time model holding for SCs of any dimension, we reveal that fixing those inconsistencies puts a topological condition on the interaction structure the model can describe, namely that two group interactions can only share one node. In order to satisfy this constraint, we introduce the notion of edge-disjoint edge clique cover (EECC) of a SC and a heuristic to find it. By means of our Microscopic Epidemic Clique Equations (MECLE), we provide a two-fold extension of the existent discrete-time complex contagion models, accounting for higher-order correlations and group interactions (if any) at the same time. We prove those dynamical correlations to be essential to describe how the critical point depends on the higher-order couplings. Lastly, different approaches to treat group interactions sharing multiple nodes are also discussed, while hinting at easy adaptations of ours.

\section*{\large{Results and Discussion}}
\label{sec:Results}

\noindent\textbf{The over-counting problem.} Let us first introduce all the basic notions needed to state the problem. We start from the interaction structures we used, simplicial complexes. A simplicial complex ${\mathcal K}$ is a subset of the power-set $2^V$ of a vertex set $V$, endowed with the hereditary property: given $f\in{\mathcal K}$ and $f^\prime\subseteq f$, then $f^\prime\in{\mathcal K}$. Note that we can neglect the empty set from $2^V$, for it does not have any practical interest here. The elements of ${\mathcal K}$ are called faces, and a $n$-dimensional face (or $n$-face) is a subset of $V$ made of $n+1$ nodes. Given a face $f$, its power-set $2^f$ is called a simplex. If $f$ is a $n$-face, $2^f$ is a $n$-simplex. Giving a geometrical interpretation to ${\mathcal K}$, a $n$-simplex is also the $n$-dimensional polytope being the convex closure of its $n+1$ vertices. For example, given $V=\{i,j,k\}$ and ${\mathcal K}=2^V$, $\{i\}$ is a $0$-simplex (a point), $\{\{i,j\},\{i\},\{j\}\}$ is a $1$-simplex (a segment), $\{\{i,j,k\},\{i,j\},\{i,k\},\{j,k\},\{i\},\{j\},\{k\}\}$ is a $2$-simplex (a triangle).

If a simplex in ${\mathcal K}$ is not included in any other simplex in ${\mathcal K}$, then it is said to be maximal. If $d$ is the maximum dimension of the faces in ${\mathcal K}$, then ${\mathcal K}$ is $d$-dimensional and is called simplicial $d$-complex. The underlying graph ${\mathcal K}^{(1)}$ of ${\mathcal K}$ is the graph induced by the $1$-simplices in ${\mathcal K}$, i.e., the graph whose node set is $V$ and whose edge set is the set of all the $1$-faces in ${\mathcal K}$. Consequently, a $n$-simplex induces a $(n+1)$-clique, made of $\binom{n+1}{2}$ edges (i.e., $1$-faces), in ${\mathcal K}^{(1)}$. If a clique is not part of a larger one, it is said to be maximal. Finally, ${\mathcal K}$ is said to be $q$-connected if, given any two simplices $s_1,s_2\subset{\mathcal K}$, there exists a sequence of simplices connecting $s_1$ and $s_2$ such that any two adjacent simplices of the sequence have (at least) a $q$-face in common; if ${\mathcal K}$ is connected, then ${\mathcal K}^{(1)}$ is $0$-connected ---as any other connected graph.

To identify whether a $n$-clique $c$ in ${\mathcal K}^{(1)}$ corresponds to a $(n-1)$-simplex in ${\mathcal K}$ or it is just a $n$-clique also in ${\mathcal K}$, we introduce a binary variable $g$, the group classifier, defined to be $1$ or $0$, respectively; by which $c$ is regarded as a $(g,n)$-clique. Since a $1$-simplex is equivalent to a $2$-clique, we choose to assign $g=0$ to any $2$-clique, so that $n\geqslant 3$ when $g=1$. From now on, unless explicitly specified, we refer to the $g$-classified cliques in ${\mathcal K}^{(1)}$ simply as `cliques'. These are the building blocks of our description.

Going to the dynamics of interest, we adopt, without loss of generality, a standard epidemiological terminology. We consider a discrete-time susceptible-infected-susceptible (SIS) dynamics on a SC. Let $\beta^{(n)}$ be the probability with which a susceptible node in a group of $n+1$ nodes (a $n$-face of the SC) gets infected when all the other $n$ nodes are infected; and $\mu$ be the probability with which an infected node recovers. Due to the hereditary property, if $n$ nodes are infected, and thus can pass the infection as a group of $n$ nodes, then also any subset of that group can: for any $k\leqslant n$, there are $\binom{n}{k}$ subsets passing the infection with probability $\beta^{(k)}$ as a group of $k$ nodes. In other words, to each state configuration of the $n$ nodes (i.e., which of them are infected) corresponds a set of concurrent channels of infections, and these channels are correlated. Compared to existing models \cite{matamalas2020abrupt}, we take a first step forward by accounting for such correlation, allowing us to compute the critical point. Then, when computing the probability for a node to get infected within a group, the contribution coming from each state configuration of the group consists of a product over the concurrent channels that configuration admits. Additionally, each contribution is weighted by the probability for the group to be in that configuration.

Now, given a clique, no matter whether it conveys ($g=1$) or not ($g=0$) group interactions, we want to account also for the dynamical correlations among the states of the nodes in it. However, if $s$ of such cliques have $m\geqslant 2$ nodes in common, the contribution to the infection of one of the $m$ nodes, coming from the others $m-1$, is counted $s$ times instead of once. This is because the $m-1$ nodes would appear in the state configuration probabilities associated with each of the $s$ cliques. To note that if the cliques are just edges, then necessarily $m=1$, meaning that the over-counting is excluded in pair-wise models.

\vspace{4.ex}
\noindent\textbf{Edge-disjoint Edge Clique Cover.} To avoid the over-counting, we aim at covering all the edges of ${\mathcal K}^{(1)}$ by means of a set of cliques such that any two of them share at most one node, giving what we call here an edge-disjoint edge clique cover. Cliques sharing more than one node are consequently decomposed in lower-order cliques. This comes with no essential repercussions when the decomposed cliques have all $g=0$. Otherwise, some group interactions would be ignored, implying the model to strictly apply for group interactions sharing at most one node. Interestingly, as shown later on, the model remains reliable when the over-counted interactions are relatively scarce.

\begin{figure}[tb!]
  \centering
    \includegraphics[width=.99\linewidth]{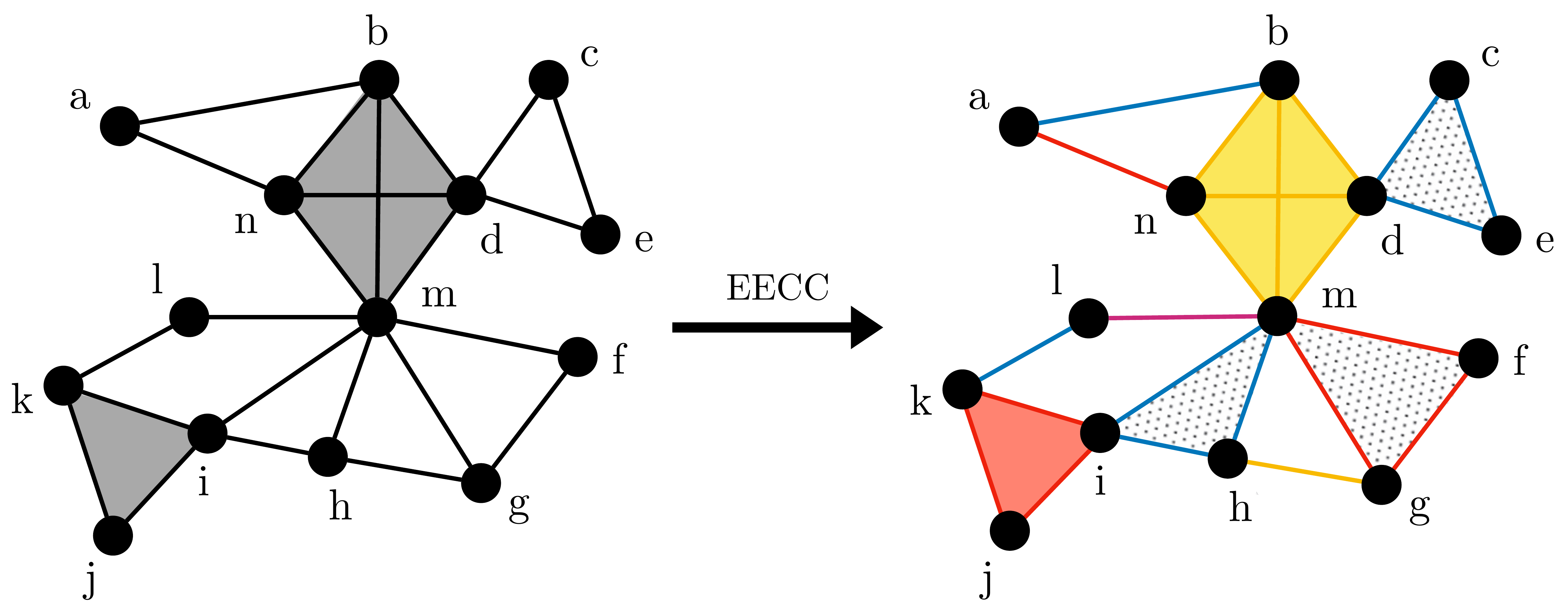}

	\caption{\fontfamily{cmss}\selectfont{\textbf{Edge-disjoint edge clique cover (EECC) of a small simplicial complex (SC).} The SC, shown on the left, consists of fourteen nodes ($0$-simplices), identified via letters, connected by one $3$-simplex, one $2$-simplex, and fourteen $1$-simplices. Grey areas indicate $r$-simplices with $r\geqslant2$, including $r+1$ nodes each. The EECC of the SC is shown on the right, where colored and dotted areas are used to visualize, respectively, the $(1,r)$-cliques and $(0,r)$-cliques in it, with colors carrying no specific meanings. The SC is decomposed in: one $(1,4)$-clique, $\{b,d,m,n\}$; one $(1,3)$-clique, $\{i,j,k\}$; three $(0,3)$-cliques, $\{c,d,e\}$, $\{f,g,m\}$, $\{h,i,m\}$; and five $(0,2)$-cliques, $\{a,b\}$, $\{a,n\}$, $\{g,h\}$, $\{k,l\}$, $\{l,m\}$. The underlying subgraph induced by the subset $\{a,b,d,m,n\}$ originally consists of a $(1,4)$-clique and a $(0,3)$-clique. To preserve the group interaction mediated by the $(1,4)$-clique, is preferable to include this in the EECC and then break the $(0,3)$-clique into two (non-maximal) $(0,2)$-cliques. Besides, the underlying subgraph induced by the subset $\{f,g,h,i,m\}$ is made of three overlapping $(0,3)$-cliques, and the EECC is in this case obtained by including $\{f,g,m\}$ and $\{h,i,m\}$ (and then the remaining edge $\{g,h\}$), instead of $\{g,h,m\}$ first.}
	}
  \label{fig:clique_decom}
\end{figure}

Covering the structure by cliques only, we are able to give a unique equation holding for any order~$n$; an impossible task via generic, less symmetric substructures. Furthermore, since we want to capture as much correlations and group interactions as possible, we want the cover to consist of the least possible number of cliques. Finding such minimal set of edge-disjoint cliques is closely related to the edge clique cover problem, known to be NP-complete \cite{kou1978covering}. Heuristics are thus necessary to estimate the solution in large graphs. For convenience, from now on, we reserve the acronym EECC to sets which are solutions of the problem.

If all the maximal cliques in ${\mathcal K}^{(1)}$ are edge-disjoint, then ${\mathcal K}^{(1)}$ admits a unique EECC, simply given by the set of the maximal cliques in it. Otherwise, ${\mathcal K}^{(1)}$ generally admits multiple EECCs. See Fig.~\ref{fig:clique_decom} for illustration.

To estimate an EECC, we propose the following greedy heuristic (the best among several options we conceived for this task). Given a graph $G$, the heuristic proceeds as follows:


{\small
\begin{enumerate}
\itemsep0.0em
  \item Find the set $C$ of all the maximal cliques in $G$
  \item Include in the EECC and remove from $C$ all the elements in $C$ that do not share edges with other elements in $C$
  \item While $C$ is not empty
  \begin{enumerate}
    \item For every maximal clique $c\in C$, compute the score $r_{c}$, defined as the fraction of edges that $c$ shares with the other elements in $C$
    \item Consider the elements of $C$ with the lowest score; include in the EECC and remove from $C$ (one, randomly chosen, of) the element(s) of highest order among them
  \end{enumerate}
\end{enumerate}
}

\noindent Noteworthy, when dealing with highly modular structures ---as those representing many real social systems--- in which communities of nodes are loosely connected among them, the search for an (edge-disjoint) edge clique cover can be speeded up. Indeed, each time two regions of the structure are joined by bridging cliques (which are evidently maximal), the problem of finding the optimal cover of the whole structure reduces to that of finding it in each of the two, smaller regions.

Now, whenever the maximal cliques in ${\mathcal K}^{(1)}$ are not all edge-disjoint, some cliques are forced to be decomposed in sub-cliques during step 4. Since decomposing a $(1, \cdot)$-clique in ($0$-connected) sub-cliques means also neglecting some group interactions, whenever a $(0,\cdot)$-clique and a $(1,\cdot)$-clique are not edge-disjoint, we prefer to include the latter in the EECC. This additional difficulty disappears whenever the SC is $0$-connected or when it has dimension~$1$ (i.e., when it is a graph). Thus, being ${\mathcal K}$ a SC and ${\mathcal K}^{(1)}$ its $g$-classified underlying graph, a EECC of ${\mathcal K}$ is constructed as follows:

{\small
\begin{enumerate}[I.]
\itemsep0.0em
  \item Consider the subgraph $G_1\subseteq{\mathcal K}^{(1)}$ induced by the nodes in the maximal $(1, \cdot)$-cliques in ${\mathcal K}$. Find a EECC of $G_1$; let us call ${\mathcal S({\mathcal K})}$ the resulting EECC
  \item Consider the subgraph $G_0={\mathcal K}^{(1)}\backslash G_1$. Find a EECC of $G_0$; let us call it ${\mathcal C({\mathcal K})}$
  \item ${\mathcal D({\mathcal K})}\equiv({\mathcal S({\mathcal K})},{\mathcal C({\mathcal K})})$ is the estimated EECC of ${\mathcal K}$
\end{enumerate}
}

A basic question is about the dependence of the prediction made by the model when different EECCs are estimated for a given structure. Indeed, while the effectiveness of the heuristic ensures a better performance of the model, its robustness is an indispensable quality, as we look for a reliable model giving certain results when fed up with a certain structure, of which a minimal EECC is estimated. In Supplementary Note~1 and Supplementary Figs.~1 and~2, our model is shown to be robust under EECC variability. Therefore, given a SC, it is safe to make use of the first EECC computed for it.

Calling $m_1$ the maximum order of the cliques we want to include in ${\mathcal S({\mathcal K})}$ (i.e., considering simplices of dimension up to $m_1-1$ in ${\mathcal K}$), $m_1$ could be smaller than $\omega_1$, the maximum order of the cliques in $G_1$. In such case, when looking for an EECC of $G_1$, those maximal cliques in $G_1$ of order greater than $m_1$ must be decomposed in edge-disjoint sub-cliques (corresponding to sub-simplices in ${\mathcal K}$) of variable order $m^\prime\in\{1,\dots ,m_1\}$. Clearly, the higher is $m_1$, the higher is the order of the group interactions (and of the correlations within them) included in the description. Overall, as long as the proportion of obviated group interactions is small enough, the deviations from the complete dynamics are comparatively negligible; or alternatively, the error made by including non-$0$-connected simplices is negligible.

In the building of an EECC, different values $m_0\geqslant 2$ for the maximum order to be considered for $(0, \cdot)$-cliques can also be chosen. The higher is $m_0$, the higher is the order of the captured dynamical correlations within the cliques in ${\mathcal K}$. In any case, $m_0\leqslant\omega_0$, being $\omega_0$ the maximum order of the cliques in $G_0$.

Summarizing, the couple $(m_0,m_1)$ identifies the considered implementation of the MECLE.


\vspace{4.ex}
\noindent\textbf{Microscopic Epidemic Clique Equations.} Given a $(g,n)$-clique $\{i_1,\dots,i_n\}$ in ${\mathcal D({\mathcal K})}$, with $n \leqslant m\equiv\textup{max}\{m_0,m_1\}$, we indicate with $P_{i_1\dots i_n,g}^{\sigma_{i_1} \dots \sigma_{i_n}}$ the joint probability that node $i_1$ is in the state $\sigma_{i_1}$, node $i_2$ is in the state $\sigma_{i_2}$, etc., where $\{\sigma_{i_1},\dots,\sigma_{i_n}\} \in \{S,I\}^n$. Besides, with $P_{i_1\dots i_{k-1} i_{k+1}\dots i_n | i_k,g}^{\sigma_{i_1} \dots\sigma_{i_{k-1}}\sigma_{i_{k+1}}\dots \sigma_{i_n}|\sigma_{i_k}}$ we indicate the conditional probability that nodes $i_1,\dots,i_{k-1},i_{k+1},\dots,i_n$ are in their respective states $\sigma_{i_1} \dots\sigma_{i_{k-1}}\sigma_{i_{k+1}}\dots\sigma_{i_n}$, given that node $i_k$ is in the state $\sigma_{i_k}$. Clearly, the normalization condition must hold:
\begin{equation}
  \sum_{\{\sigma_{i_k}\}_{k=1,\dots,n}} P_{i_1\dots i_n,g}^{\sigma_{i_1} \dots \sigma_{i_n}}=1
\end{equation}

Indicating with $\{\sigma_{i_k}\}_{k=1,\dots,n}$ the states at time $t$ and with $\{\sigma_{i_k}^\prime\}_{k=1,\dots,n}$ those at time $t+1$, the MECLE model dynamic equation governing the evolution of the state of a $(g,n)$-clique $\{i_1,\dots,i_n\}$ reads
\begin{align}
  \notag  P&_{i_1\dots i_n,g}^{\sigma_{i_1}^\prime \dots \sigma_{i_n}^\prime}\left(t+1\right)
  \\
  &=\sum_{\{\sigma_{i_1}, \dots, \sigma_{i_n}\}} P_{i_1\dots i_n,g}^{\sigma_{i_1} \dots \sigma_{i_n}}\left(t\right)~\Phi_g\left(\{\sigma_{i_k}\},\{\sigma_{i_k}^\prime\};\{\beta^{(s)}\},\mu\right)
  \label{eq:DE_n_clique}
\end{align}

\noindent where
\begin{align}
  \notag \Phi_g\left(\{\sigma_{i_k}\},\{\sigma_{i_k}^\prime\};\{\beta^{(s)}\},\mu\right) &
  \\
  =\prod_{k=1}^n \phi_{i_k,g}&\left(\{\sigma_{i_k}\},\sigma_{i_k}^\prime;\{\beta^{(s)}\},\mu\right)
  \label{eq:Phi_all}
\end{align}

\noindent is the transition probability from the starting state $\{\sigma_{i_k}\}_{k=1,\dots,n}$ to the arrival state $\{\sigma_{i_k}^\prime\}_{k=1,\dots,n}$. It is understood that $\Phi_g$ is computed at time $t$. This is expressed as a product over the single-node transition probabilities $\{\phi_{i_k,g}\}$, given by
\begin{align}
  \notag \phi_{i_k,g}&\left(\{\sigma_{i_k}\},\sigma_{i_k}^\prime;\{\beta^{(s)}\},\mu\right)
  \\
  \notag &={\mathbb{1}}[\sigma_{i_k}=I, \sigma_{i_k}^\prime=I]\left(1-\mu\right)
  \\
  \notag &+{\mathbb{1}}[\sigma_{i_k}=I, \sigma_{i_k}^\prime=S]~\mu
  \\
  \notag &+{\mathbb{1}}[\sigma_{i_k}=S, \sigma_{i_k}^\prime=I]\left[1-\frac{w_{N_I,g}^{(n-1)}}{q_{i_k(\neg i_k),g}^{(n-1)}}\prod_{(g^\prime,r)} q_{i_k,g^\prime}^{(r)}\right]
  \\
  &+{\mathbb{1}}[\sigma_{i_k}=S, \sigma_{i_k}^\prime=S]\left[\frac{w_{N_I,g}^{(n-1)}}{q_{i_k(\neg i_k),g}^{(n-1)}}\prod_{(g^\prime,r)} q_{i_k,g^\prime}^{(r)}\right]
  \label{eq:phi_ik_g}
\end{align}

\noindent where ${\mathbb{1}}[p]$ gives $1$ if condition $p$ is fulfilled and $0$ otherwise; $N_I=\left\vert\{i_{k=1,\dots ,n}\vert\sigma_{i_k}=I\}\right\vert$ is the number of infected nodes in the starting state; $w_{N_I,g}^{(n-1)}\equiv w_{N_I,g}^{(n-1)}\left(\{\beta^{(s)}\}\right)$ is the probability that a susceptible node ($i_k$) does not get infected within a $(g,n)$-clique ($\{i_1,\dots ,i_k,\dots, i_n\}$) whose state configuration ($N_I\leqslant n-1$) is known, reading
\begin{align}
  \notag w_{N_I,g}^{(n-1)} =& ~{\mathbb{1}}[g=0]\left(1-\beta^{(1)}\right)^{N_I}
  \\
  +&~{\mathbb{1}}[g=1]\prod_{s=1}^{N_I}\left(1-\beta^{(s)}\right)^{\binom{N_I}{s}}
  \label{eq:w}
\end{align}

\noindent and $q_{i_k,g^\prime}^{(r)}\equiv q_{i_k,g^\prime}^{(r)}\left(\{\beta^{(s)}\}\right)$ is the probability that node $i_k$ does not get infected via any of the $(g^\prime,r+1)$-cliques incident on it, that is
\begin{align}
  \notag q_{i_k,g^\prime}^{(r)} = &\prod_{\scriptstyle {\begin{array}{c}{\{j_1,\dots,j_r\}}\\ \in\Gamma_{i_k,g^\prime}^{(r)}\end{array}}} \left[1-\sum_{l=1}^r\frac{1-w_{l,g^\prime}^{(r)}}{l!\left(r-l\right)!}\right.
  \\
  &\hspace{5pt}\left.\times\sum_{k_1\neq\dots\neq k_r=1}^r P_{j_{k_1}\dots j_{k_l}j_{k_{l+1}}\dots j_{k_r}|i_k,g^\prime}^{I\dots IS\dots S|S}\right]
  \label{eq:q}
\end{align}

\noindent where $\Gamma_{i_k,g^\prime}^{(r)}$ indicates the set of $r$-tuples of indexes corresponding to subsets of $r$ nodes forming a $(g^\prime,r+1)$-clique with $i_k$, and $q_{i_k(\neg i_k),g}^{(n-1)}$ coincides with $q_{i_k,g}^{(n-1)}$ expect for excluding the considered $(g,n)$-clique $\{i_1,\dots ,i_k,\dots, i_n\}$ from the product. Finally, the products in Eq.~(\ref{eq:phi_ik_g}) are performed over the couples $\{(g^\prime ,r)\}$ such that $2\leqslant r\leqslant m_0$ for $g^\prime=0$, and $3\leqslant r\leqslant m_1$ for $g^\prime=1$.

In Eq.~(\ref{eq:Phi_all}) the single-node transition probabilities, $\phi$, are treated as independent from each other within the time step. This merely derives from the implicit assumption that all the events within a given time step are simultaneous, and therefore not causally related. Simply put, the state of a node at time step $t+1$ only depends on its state and on the states of its neighbors at the previous time step $t$, as in any markovian model.

Importantly, to get the expression for $q_{i_k,g^\prime}^{(r)}$ we have adopted the following closure
\begin{align}
  \notag P_{j_{k_1}\dots j_{k_l}j_{k_{l+1}}\dots j_{k_r}i_1\dots i_k\dots i_n}^{I\dots IS\dots S\sigma_{i_1}\dots S\dots\sigma_{i_n}} \hspace{100pt} &
  \\
  = \frac{P_{j_{k_1}\dots j_{k_l}j_{k_{l+1}}\dots j_{k_r}i_k,g^\prime}^{I\dots IS\dots SS} P_{i_1\dots i_k\dots i_n,g}^{\sigma_{i_1}\dots S\dots\sigma_{i_n}}}{P_{i_k}^S}&
  \label{eq:closure}
\end{align}

\noindent (the classifier $g$ is assigned only to the cliques in ${\mathcal D({\mathcal K})}$). Therefore, by definition of conditional probability, the probability $P_{j_{k_1}\dots j_{k_l}j_{k_{l+1}}\dots j_{k_r}|i_1\dots i_k\dots i_n}^{I\dots IS\dots S|\sigma_{i_1}\dots S\dots\sigma_{i_n}}$ appears in Eq.~(\ref{eq:q}) as $P_{j_{k_1}\dots j_{k_l}j_{k_{l+1}}\dots j_{k_r}|i_k,g^\prime}^{I\dots IS\dots S|S}$. Intuitively, each clique in ${\mathcal D({\mathcal K})}$ is treated as an independent dynamical unit of the system, accounting for the correlations among the states of the nodes it includes. In the form of a generalization of the classical pair approximation \cite{henkel2008non}, the dynamical correlations between two adjacent cliques are conveyed by the state probability of the node they share (in the denominator to avoid double counting). In this regard, the larger the adjacent cliques, the smaller the expected influence of the state of the shared node on all the other ones. For this reason, the presented closure ---and consequently the MECLE--- is expected to gain further accuracy with the increasing of the order of the cliques.

The probability $P_i^I$ for the single node $i$ of being infected is computed as a marginal probability from any $(g,n)$-clique including $i$, as
\begin{align}
  \notag P_i^I\left(t+1\right) &= \sum_{\{\sigma_{\neg i}\}} P_{i\{\neg i\},g}^{I\{\sigma_{\neg i}\}}\left(t+1\right)&
  \\
  &= P_i^I\left(t\right)\left(1-\mu\right)+P_i^S\left(t\right)\left(1-\prod_{(g,r)}q_{i,g}^{(r)}\right)
  \label{eq:P_i}
\end{align}

\noindent in which $\{\neg i\}$ and $\{\sigma_{\neg i}\}$ indicate, respectively, the set of the other $n-1$ nodes in the clique and their states. Equation~(\ref{eq:P_i}) is, consistently, also found taking $n=1$ in Eq.~(\ref{eq:DE_n_clique}) ($w_{N_I,g}^{(0)}=1$, $q_{i,g}^{(0)}=1$). The average value $\rho$ of the single node probabilities,
\begin{equation}
  \rho\left(t\right)=\frac{1}{N}\sum_{i=1}^N P_i^I\left(t\right)
  \label{eq:rho}
\end{equation}
is the epidemic prevalence, which is also the order parameter of the system.

It is important to note that the used closure, Eq.~(\ref{eq:closure}), apart from preserving the $n$-node state correlations of a considered $(g,n)$-clique, is the only one making feasible the marginalization of Eq.~(\ref{eq:DE_n_clique}) to get Eqs.~(\ref{eq:P_i})-(\ref{eq:rho}) and, consequently, get the expression for the epidemic threshold, through Eqs.~(\ref{eq:eigen}) to~(\ref{eq:epi_th}) in Methods.

At any time step $t$, Eq.~(\ref{eq:P_i}) yields one constraint for each of the $n$ nodes in a $(\cdot ,n)$-clique, leaving $2^n-n-1$ independent state probabilities to be determined, one being fixed by the normalization. Therefore, if ${\mathcal K}$ has $N$ nodes and ${\mathcal D}({\mathcal K})$ consists of $C^{(n)}$ $(\cdot ,n)$-cliques, $n \leqslant m$, the MECLE is defined by a system of $N+\sum_{n=2}^m \left(2^n-n-1\right)C^{(n)}$ independent equations.

To help in the understanding of the model, we show in Supplementary Note~2 all the equations for the particular case of the simplicial $2$-complexes, i.e., the $(3,3)$ implementation of the MECLE.

Finally, we can frame the existent models in the $(m_0,m_1)$ notation. As $m_1=2$ implies ${\mathcal S({\mathcal K})}=\varnothing$, $\{(m_0,2)\}_{2\leqslant m_0\leqslant\omega_0}$ is the class of models accounting for correlations within cliques of order up to $m_0$ on graphs (simple contagion). In particular, $(2,2)$ gives the Epidemic Link Equations (ELE) model \cite{matamalas2018effective}. The Microscopic Markov Chain Approach (MMCA) \cite{gomez2010discrete} is then recovered from $(2,2)$ by assuming that $P_{ji}^{IS}=P_j^I P_i^S$ (i.e., $P_{j|i}^{I|S}=P_j^I$ in Eq.~(\ref{eq:q})). Considering group interactions, the simplicial ELE and MMCA \cite{matamalas2020abrupt} fall outside the MECLE class of models, for they do not account for the correlations among the concurrent channels of infection within simplices. This, alongside the consideration of higher-order dynamical correlations among nodes' states, is precisely the refinement made here.

\vspace{4.ex}
\noindent\textbf{Results for simplicial $\boldsymbol{2}$-complexes.} We apply here the developed formalism to the case of simplicial $2$-complexes. As we are only using synthetic structures, the SCs are constructed from some graph (the future underlying graph) converting into $(1,3)$-cliques (i.e., $2$-faces) a fraction $p_\triangle$ of the $3$-cliques allowed by the EECC of the graph, while considering as $(0,3)$-cliques the remaining fraction $1-p_\triangle$. Specifically, if all the $3$-cliques are converted ($p_\triangle=1.0$), the resulting SCs are the clique complexes of the respective original graphs. To better appreciate the improvement made here, we mainly show results for clique complexes, although notable improvements can be generally found for any value of $p_\triangle$ depending on the used structure. Conveniently, if $p_\triangle$ is not specified then the structure is understood to be a clique complex. We identify a such generated SC adding `SC' to the name of the graph model used for it: a `Dorogovtsev-Mendes SC', for instance, is built upon a graph given by the Dorogovtsev-Mendes generative model \cite{dorogovtsev2001size}. Specifically, the `random SC' is obtained as follows: first we generate a simplicial $2$-complex through the random simplicial complex model \cite{iacopini2019simplicial}; then we consider its underlying graph and compute its EECC; finally, the random SC is got, as before, by converting $3$-cliques in $2$-faces, in this way ensuring the SC to be $0$-connected.

\begin{figure*}[t!]
  \centering
  \includegraphics[width=.98\linewidth]{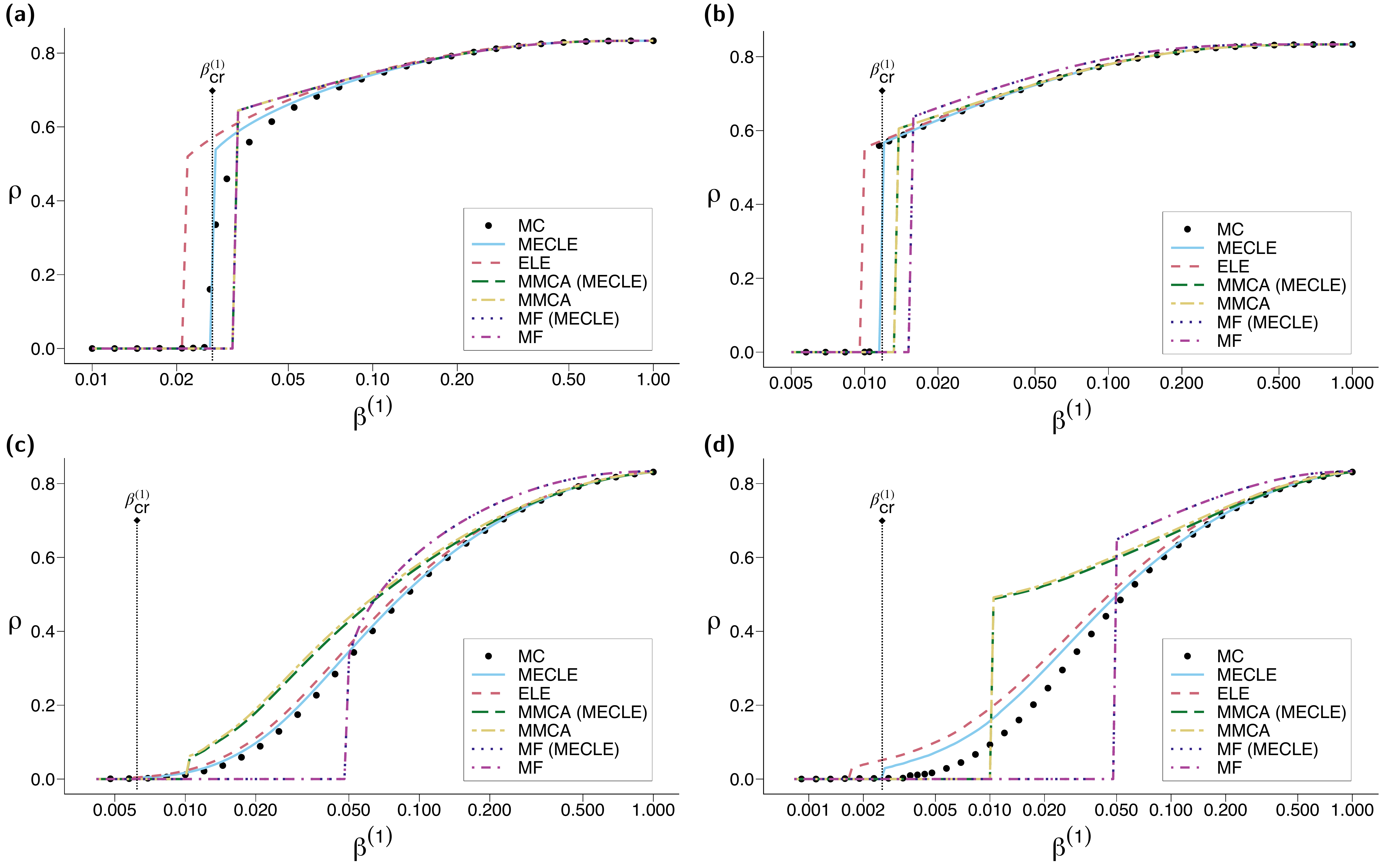}

  \caption{\fontfamily{cmss}\selectfont{\textbf{Epidemic prevalence {\em{\textrho}} as function of the edge infection probability {\em{\textbeta}}$\boldsymbol{^{(1)}}$ on different $\boldsymbol{2}$-dimensional simplicial complexes (SCs).} Results obtained from Monte Carlo (MC) simulations are depicted by dots, while lines represent the analytically computed prevalence using the indicated models. MMCA and MMCA(MECLE) refer to the Microscopic Markov Chain approximation of, respectively, the simplicial Epidemic Link Equations (ELE) model and the Microscopic Epidemic Clique Equations (MECLE) model, as obtained by considering the state probabilities of the nodes as uncorrelated; while MF and MF(MECLE) refer to their homogeneous mean-field approximations (see Methods). Note that MF and MF(MECLE) are indistinguishable at the used scale. The value of the epidemic threshold, as computed in the MECLE through Eqs.~(\ref{eq:eigen}) to~(\ref{eq:epi_th}) in Methods, is marked with a vertical dotted line. The recovery probability is fixed to $\mu=0.2$. (a)~Periodic triangular SC with $\bar{k}^{(0,1)}=0.00$, $\bar{k}^{(0,2)}=0.00$ and $\bar{k}^{(1,2)}=3.00$, being $\bar{k}^{(g,r)}$ the mean number of $(g,n+1)$-cliques incident on a node, and triangle infection probability $\beta^{(2)}=0.25$; the relative error in locating the epidemic threshold is $\varepsilon_{\beta^{(1)}}\approx 0.08$ for MECLE and $\varepsilon_{\beta^{(1)}}\approx 0.12$ for ELE. (b)~Random SC with $\bar{k}^{(0,1)}=4.10$, $\bar{k}^{(0,2)}=0.00$ and $\bar{k}^{(1,2)}=3.95$, and $\beta^{(2)}=0.15$; $\varepsilon_{\beta^{(1)}}\approx 0.06$ for MECLE and $\varepsilon_{\beta^{(1)}}\approx 0.09$ for ELE. (c)~Dorogovtsev-Mendes SC with $\bar{k}^{(0,1)}=1.10$, $\bar{k}^{(0,2)}=0.00$ and $\bar{k}^{(1,2)}=1.45$, and $\beta^{(2)}=0.25$; $\varepsilon_{\beta^{(1)}}\approx 0.07$ for MECLE and $\varepsilon_{\beta^{(1)}}\approx 0.16$ for ELE. (d)~Same as (c) but with $\beta^{(2)}=0.50$; $\varepsilon_{\beta^{(1)}}\approx 0.23$ for MECLE and $\varepsilon_{\beta^{(1)}}\approx 0.50$ for ELE.}}
  \label{fig:results}
\end{figure*}

In Figure~\ref{fig:results} we compare the prevalence $\rho$ obtained using Monte Carlo (MC) simulations (see Methods for details), the MECLE model, and the other discrete-time markovian models, i.e., the simplicial ELE and MMCA models \cite{matamalas2020abrupt}, for different $0$-connected clique complexes. For all structures, the improvement brought by the MECLE with respect to the simplicial ELE is substantial, in both predicting the epidemic prevalence and, even more, locating the critical point, for which the relative errors $\varepsilon_{\beta^{(1)}}$ are reported. Note that the predictions is expected to improve for increasing order of the cliques, since the precision of the used closure, Eq.~(\ref{eq:closure}), grows with it as well, especially in the case of populations arranged in lowly inter-connected dense communities.

\begin{figure*}[tb!]
  \centering
  \includegraphics[width=.98\linewidth]{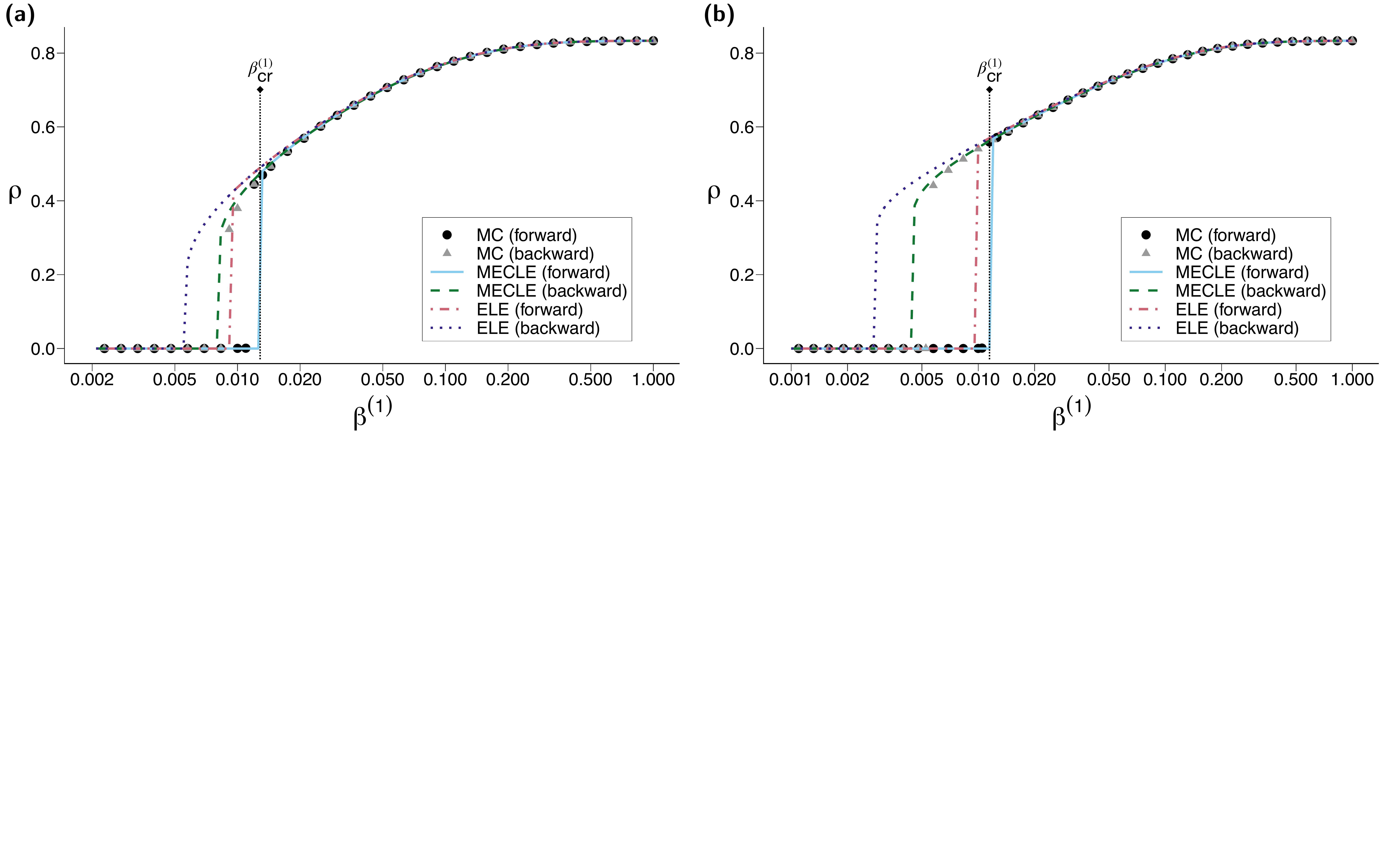}

  \caption{\fontfamily{cmss}\selectfont{\textbf{Hysteresis cycle of the epidemic prevalence {\em{\textrho}} with respect to the edge infection probability {\em{\textbeta}}$\boldsymbol{^{(1)}}$ on different $\boldsymbol{2}$-dimensional simplicial complexes (SCs).} Results obtained from Monte Carlo (MC) simulations are depicted by dots, while lines represent the analytically computed prevalence using the simplicial Epidemic Link Equations (ELE) model and the Microscopic Epidemic Clique Equations (MECLE) model. The `forward' and `backward' curves are obtained through small equilibrium transformations taking as initial value $\rho_0$ the equilibrium value of $\rho$ got at the next smaller and next greater value of $\beta^{(1)}$, respectively. The hysteresis cycle reveals the bi-stable region. The computed value of the (forward) epidemic threshold is marked with a vertical dotted line. The recovery probability is fixed to $\mu=0.2$. (a)~Random SC with $\bar{k}^{(0,1)}=4.10$,  $\bar{k}^{(0,2)}=1.58$ and  $\bar{k}^{(1,2)}=2.37$ ($p_\triangle = 0.6$), being $\bar{k}^{(g,r)}$ the mean number of $(g,n+1)$-cliques incident on a node, and triangle infection probability $\beta^{(2)}=0.25$; the relative errors in locating the epidemic threshold are $\varepsilon_{\beta^{(1)}}^{\text{for.}}\approx 0.14$ and $\varepsilon_{\beta^{(1)}}^{\text{back.}}\approx 0.10$ for MECLE, and $\varepsilon_{\beta^{(1)}}^{\text{for.}}\approx 0.18$ and $\varepsilon_{\beta^{(1)}}^{\text{back.}}\approx 0.38$ for ELE. (b)~Random SC with $\bar{k}^{(0,1)}=4.10$, $\bar{k}^{(0,2)}=0.00$ and $\bar{k}^{(1,2)}=3.95$ ($p_\triangle =1.0$), and $\beta^{(2)}=0.15$; $\varepsilon_{\beta^{(1)}}^{\text{for.}}\approx 0.06$ and $\varepsilon_{\beta^{(1)}}^{\text{back.}}\approx 0.15$ for MECLE, and $\varepsilon_{\beta^{(1)}}^{\text{for.}}\approx 0.09$ and $\varepsilon_{\beta^{(1)}}^{\text{back.}}\approx 0.47$ for ELE.}}
  \label{fig:results_hys}
\end{figure*}

We then leverage the prominent jump of the discontinuous transition found for random SCs, to illustrate, in Fig.~\ref{fig:results_hys}, the existence of a hysteresis cycle, enclosing a bi-stable region. This is in line with recent findings \cite{iacopini2019simplicial,landry2020effect,bodo2016sis,jhun2019simplicial,de2020social,matamalas2020abrupt}. Again, the simplicial ELE is outperformed by the MECLE, especially in predicting the `backward' curves, as the overestimation made by the former (see next subsection) is emphasized in that case.

\begin{figure}[tb!]
  \centering
  \includegraphics[width=.99\linewidth]{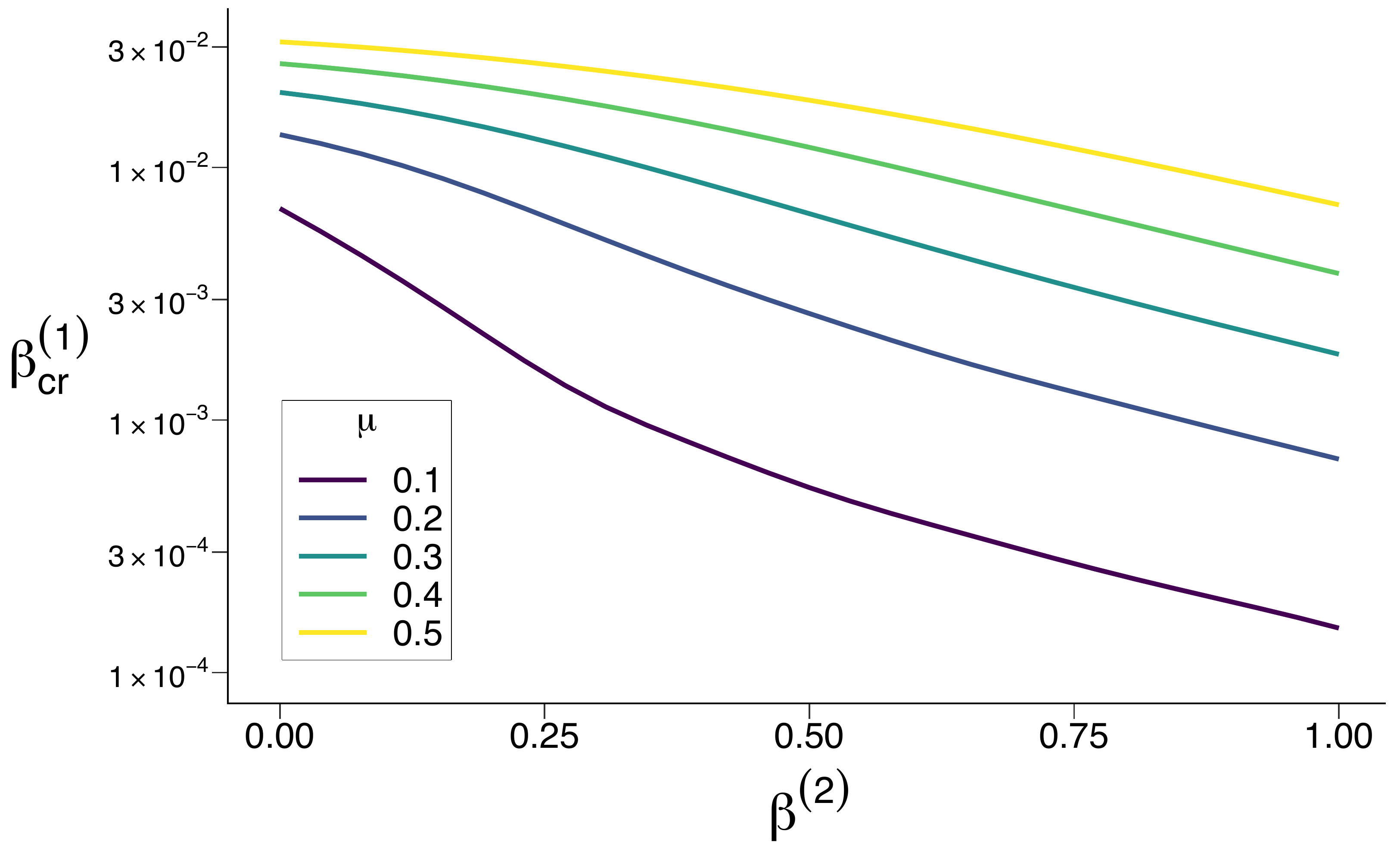}

  \caption{\fontfamily{cmss}\selectfont{\textbf{Dependence of the epidemic threshold on the triangle infection probability {\em{\textbeta}}$\boldsymbol{^{(2)}}$.} $\beta^{(1)}_{\textup{cr}}$, computed via Eqs.~(\ref{eq:eigen}) to~(\ref{eq:epi_th}) in Methods, is shown against $\beta^{(2)}$ for a Dorogovtsev-Mendes simplicial complex with $\bar{k}^{(0,1)}=1.10$, $\bar{k}^{(0,2)}=0.00$ and $\bar{k}^{(1,2)}=1.45$, being $\bar{k}^{(g,r)}$ the mean number of $(g,n+1)$-cliques incident on a node. Note that Eq.~(\ref{eq:epi_th_mf}), disregarding dynamical correlations, wrongly predicts $\beta^{(1)}_{\textup{cr}}=\mu/{\bar k}=\mu/4$, $\forall\beta^{(2)}$.}}
  \label{fig:beta_cr_DM}
\end{figure}

More interestingly, our analysis clearly shows that models with uncorrelated nodes' states generally fail to pinpoint the (`forward') epidemic threshold at even a qualitative level. Indeed, by neglecting those dynamical correlations, Eq.~(\ref{eq:epi_th_mf}) (see Methods) predicts $\beta^{(1)}_{\textup{cr}}$, the value of $\beta^{(1)}$ at which the epidemic-free state becomes unstable, to be independent from the values of the higher-order infection probabilities, $\{\beta^{(s)}\}_{s>1}$. In particular, in the case of $2$-dimensional simplicial complexes, increasing enough $\beta^{(2)}$ results in the appearance of a bi-stable region, but $\beta^{(1)}_{\textup{cr}}$ does not change in those models\cite{iacopini2019simplicial,landry2020effect}. On the contrary, when the actual structure of the interactions is retained, increasing $\beta^{(2)}$ does lead $\beta^{(1)}_{\textup{cr}}$ to decrease, as arises from MC simulations and correctly predicted by the MECLE. We show this dependence in Fig.~\ref{fig:beta_cr_DM}, while it can also be grasped by comparing panels (c) and (d) in Fig.~\ref{fig:results}.

The calculation of the critical point in the MECLE (see Methods), together with the evidences coming from the MC simulations, reveal the necessity of accounting for higher-order dynamical correlations. An interaction (infection) of order $s$, with coupling (infection probability) $\beta^{(s)}$, requires $s$ nodes in a simplex to be active (infected). Therefore, in order to preserve its contribution near the critical point, the state probability with $s$ active nodes must not be neglected ---as done instead in models with uncorrelated nodes' states. In other words, near the epidemic threshold, higher-order interactions rely exclusively on comparatively higher-order correlations, so that their effect when varying the higher-order couplings, $\{\beta^{(s)}\}_{s>1}$, is observed only if such correlations are preserved. Accordingly, such dependence eludes the MF and MMCA approximations of the system. Besides, as second-order correlations within triangles are partially accounted for in the simplicial ELE, the latter predicts this dependence, but not as accurately as MECLE does. Finally, in the endemic state and away from the threshold, low-order correlations suffice to sustain higher-order interactions. Nevertheless, the higher the order of the accounted correlations, the more accurate is the quantification of the prevalence.

While a closed equation for $\beta^{(1)}_{\textup{cr}}$ is generally inaccessible for complex interaction structures, in Supplementary Note~3 and Supplementary Figs.~3 and~4, we explicitly derive the monotonous decrease of $\beta^{(1)}_{\textup{cr}}$ with respect to the higher-order infection probabilities for selected symmetrical structures. In particular, we study regular SCs, like the herein studied periodic triangular SC, and SCs built upon Friendship graphs, taken as proxies for homogeneous and heterogeneous structures, respectively. We prove that $\beta^{(1)}_{\textup{cr}}$ decays with $\beta^{(2)}$ whatever the system size $N$, and that the dependence considerably increases with the structural heterogeneity.

It should be noted that the herein observed and proved dependence of the critical point on the higher-order couplings, already showed up in previously reported numerical simulations, especially in SCs built from real data \cite{iacopini2019simplicial}. However, it went seemingly overlooked, eventually leading to claim \cite{barrat2021social} that only pair-wise interactions govern the value of $\beta^{(1)}_{\textup{cr}}$. Even though we have considered a discrete-time dynamic, instead of the continuous-time' used in those works upon which that claim is based on, we predict the qualitative shift brought by our analysis to hold in continuous-time as well. The continuous-time limit is here recovered by neglecting all those terms in the equations appearing as second or greater powers of any combination of the infection probabilities $\{\beta^{(s)}\}$ and the recovery probability $\mu$, i.e., allowing only single-node state changes. Still, the linear terms proportional to any of the infection probabilities show up in Eq.~\ref{eq:epi_th}, thus contributing to the value of the critical point. In Supplementary Note~4, we derive the continuous-time limit of the MECLE equations for simplicial $2$-complexes. Taking the example of a Dorogovtsev-Mendes SC, the predicted dependence of the critical point on $\beta^{(2)}$ is shown in Supplementary Fig.~5.

\begin{figure*}[tb!]
  \centering
  \includegraphics[width=.98\linewidth]{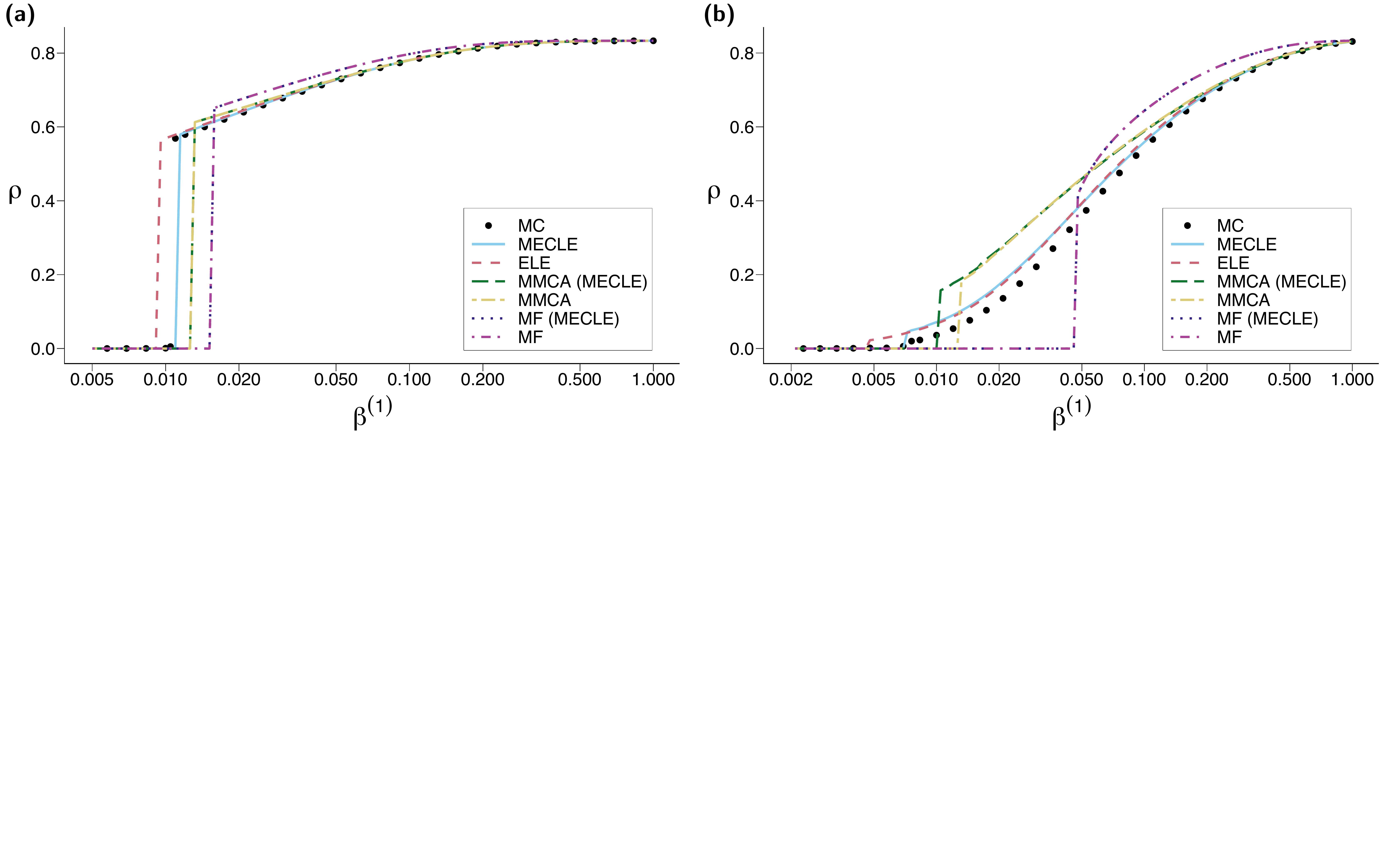}

  \caption{\fontfamily{cmss}\selectfont{\textbf{Epidemic prevalence {\em{\textrho}} as function of the link infection probability {\em{\textbeta}}$\boldsymbol{^{(1)}}$ on $\boldsymbol{2}$-dimensional simplicial complexes (SCs) with low percentage $\boldsymbol{p_\textup{s}}$ of shared non-maximal edges.} $p_\textup{s}$ is computed as the fraction of edges within $2$-faces which are included in more than one $2$-face. Besides, with $\bar{s}$ we indicate the average number of $2$-faces in which the edges corresponding to the fraction $p_\textup{s}$ are included ($\bar{s}\geqslant 2$). Results obtained from Monte Carlo (MC) simulations are depicted by dots, while lines represent the analytically computed prevalence using the indicated models. MMCA and MMCA(MECLE) refer to the Microscopic Markov Chain approximation of, respectively, the simplicial Epidemic Link Equations (ELE) model and the Microscopic Epidemic Clique Equations (MECLE) model, as obtained by considering the state probabilities of the nodes as uncorrelated; while MF and MF(MECLE) refer to their homogeneous mean-field approximations (see Methods). Note that MF and MF(MECLE) are indistinguishable at the used scale. The recovery probability is fixed to $\mu=0.2$. (a)~Random SC with $\bar{k}^{(0,1)}=2.25$, $\bar{k}^{(0,2)}=0.00$ and $\bar{k}^{(1,2)}=5.20$, being $\bar{k}^{(g,r)}$ the mean number of $(g,n+1)$-cliques incident on a node; $p_\textup{s}=0.03$, $\bar{s}=2.01$, and triangle infection probability $\beta^{(2)}=0.15$. (b)~Dorogovtsev-Mendes SC with $\bar{k}^{(0,1)}=1.10$, and $\bar{k}^{(0,2)}=0.00$ $\bar{k}^{(1,2)}=1.60$, $p_\textup{s}=0.05$, $\bar{s}=2.06$, and $\beta^{(2)}=0.25$.}}
  \label{fig:results_overlap}
\end{figure*}

Lastly, considering clique complexes with some fraction $p_\textup{s}$ of edges shared by two or more $2$-faces, we have studied how the MECLE behaves out of the bounds of the $0$-connectedness. As shown in Fig.~\ref{fig:results_overlap}, when $p_\textup{s}$ is low enough, it performs comparably to or still better than the simplicial ELE. As expected, the value of $p_\textup{s}$ above which the MECLE performs worse turns out to specifically depend on the structure, preventing us to find a simple relation. Precisely, that value resulted to be around $0.06$ for Dorogovtsev-Mendes SCs and around $0.04$ for RSCs.

\vspace{4.ex}
\noindent\textbf{Possible approaches for group interactions sharing multiple nodes.} In discrete-time models, allowing groups to share two or more nodes comes with the drawback of impeding the computation of the epidemic threshold, as shown in Methods. Forgoing the latter, let us discuss some approaches which may still come in handy.

The first option is the simplicial ELE model \cite{matamalas2020abrupt}. It describes a SIS dynamics in simplicial $2$-complexes by means of a pair approximation that, via a specific triangle closure, is able to partially account for second-order correlations. However, considering the multiple channels of infection within a simplex as mutually uncorrelated, it is not possible to marginalize the $1$-faces equations to get the single nodes one, hence neither the epidemic threshold. To further elucidate the effects of disregarding the correlations among the concurrent infections within a simplex, let us consider the toy example of a triangle upon the set of nodes $\{i,j,k\}$. In the MECLE model, the probability for $i$ to be infected by $j$ and $k$, reads
{\small
\begin{equation}
  \notag \beta^{(1)}\left(P_{jk|i}^{IS|S}+P_{jk|i}^{SI|S}\right) + \left[1-\left(1-\beta^{(1)}\right)^2\left(1-\beta^{(2)}\right)\right]P_{jk|i}^{II|S}
\end{equation}
}

\noindent In particular, taking $\beta^{(1)}=1$, the dependence on $\beta^{(2)}$ correctly disappears: no matter the infectiousness $\beta^{(2)}$ of the couple $\{j,k\}$, the probability for node $i$ to be infected equals the probability that at least one node between $j$ and $k$ is infected. On the contrary, the simplicial ELE neglects that correlation, and the above infection probability becomes
\begin{equation}
  \notag 1-\left(1-\beta^{(1)}P_{j|i}^{I|S}\right)\left(1-\beta^{(1)}P_{k|i}^{I|S}\right)\left(1-\beta^{(2)}P_{jk|i}^{II|S}\right)
\end{equation}

\noindent where $P_{jk|i}^{II|S}$ is expressed in terms of edge probabilities. Here, taking $\beta^{(1)}=1$, there is still a dependence on $\beta^{(2)}$, with the effect of both overestimating the probability of infection, hence the prevalence, and wrongly anticipating the position of the epidemic threshold. Evidently, the error grows with both $\beta^{(2)}$ and $P^{II|S}$. To notice that this analysis does not make any reference to the correlations among nodes' states a model considers. Consequently, a similar comparison holds also between the MMCA and MF approximations of the MECLE and those of the simplicial ELE, with the latter overestimating more the prevalence.

Curiously, when the $0$-connectedness is heavily broken (making the MECLE unreliable), the simplicial ELE appears to gain accuracy, especially in locating the critical point \cite{matamalas2020abrupt}. However, while we could explain the reasons why the MECLE outperforms the simplicial ELE in (nearly) $0$-connected SCs, the unexpected improvement of the latter for higher connectedness remains unclear. Changes in topological factors, like the spectral dimension (note that, independently from their geometrical dimension, $0$-connected SCs have spectral dimension very close to that of a random tree \cite{dankulov2019spectral}) or the triangles' percolation, are probably able to soften the approximations of the model. Future work addressing the role of those factors may help to better understand the limits of the simplicial ELE, while giving new, general insights about the relation between dynamical correlations and topology.

An alternative approach, assuming the considered contagion would make it usable, is to opt for more general hypergraphs by dropping the hereditary property out. One could then generalize the approach of Matamalas et al. \cite{matamalas2020abrupt} to any hypergraph, but at the aforementioned cost of accepting some inconsistencies in the marginalization of the probabilities. Besides, referring to the class of simple hypergraphs \cite{bretto2013hypergraph}, i.e., those in which a hyperedge ---what in a SC is a face--- cannot be subset of any other hyperedge, one can still resort to the fully consistent approach of the MECLE. Indeed, in such structures, to each configuration of the group corresponds a unique channel of infection, so the necessity of constraining groups to share not more than a single node can be relaxed. A model can then be easily constructed adapting Eq.~(\ref{eq:DE_n_clique}) by solely modifying the form of $w_{l,1}^{(r)}$, being $g=1$ for hyperedges including more than two nodes. For example, supposing the contagion to be maximally conservative, we would write $w_{l,1}^{(r)}={\mathbb{1}}[l=r]\left(1-\beta^{(r)}\right)$. Moreover, for linear hypergraphs \cite{bretto2013hypergraph}, in which two hyperedges can only share one node, the critical point is again calculable.

Lastly, in a regime of both high and rapid infectiousness and recovery, each set of nodes shared by a sufficient number of groups (e.g., two partners carrying out many of their activities together) could be effectively treated as a single super-node, whose state always represents that of each of the nodes it contains. This effective approach can be combined with any of the ones that have been discussed.

\end{spacing}

\vspace{4.ex}
\begingroup
\fontsize{8pt}{9pt}\selectfont

\section*{\large{Methods}}
\label{sec:methods}

\noindent{\fontfamily{cmss}\selectfont{\textbf{Epidemic threshold.}}}
Here we derive the critical point $\beta^{(1)}_{\textup{cr}}$, defined as the value of $\beta^{(1)}$ at which the inactive (epidemic-free) state becomes unstable, thus marking the onset of the active (endemic) state. In the presence of a bi-stable region, it identifies the rightmost transition.

We linearize Eq.~(\ref{eq:DE_n_clique}) by regarding of the same order $\epsilon \ll 1$, all the state probabilities containing at least one infected node, i.e., $P_{i_1\dots i_n,g}^{\sigma_{i_1} \dots \sigma_{i_n}}\sim{\cal O}(\epsilon)$ iff $\exists k: \sigma_{i_k}=I$; and consequently, $P_{i_1\dots i_n,g}^{S\dots S}\sim 1-{\cal O}(\epsilon)$. Without this assumption, higher-order dynamical correlations would be lost, and the critical point would not be correctly located. This can be interpreted as an extension of the results in Matamalas et al. \cite{matamalas2018effective}, where it is shown that $P^{II}\sim{\cal O}(\epsilon)$ for the state probability of an edge having both nodes infected.

Being interested in stationary states, the value of the time step is omitted from now on. All the states appearing in $q_{i_k,g^\prime}^{(r)}$, see Eq.~(\ref{eq:q}), include at least one infected node, therefore it takes the form,
\begin{align}
  \notag q_{i_k,g^\prime}^{(r)} = 1~ - &\sum_{\scriptstyle {\begin{array}{c}{\{j_1,\dots,j_r\}}\\ \in\Gamma_{i_k,g^\prime}^{(r)}\end{array}}}
  \left[\sum_{l=1}^r\frac{1-w_{l,g^\prime}^{(r)}}{l!\left(r-l\right)!}\right.
  \\
  &\left.\times\sum_{k_1\neq\dots\neq k_r=1}^r P_{j_{k_1}\dots j_{k_l}j_{k_{l+1}}\dots j_{k_r}i_k,g^\prime}^{I\dots IS\dots SS}\right]
  + \mathcal{O}(\epsilon^2)
  \label{eq:q_lin}
\end{align}

\noindent where the squared brackets contain ${\cal O}(\epsilon)$ terms only. It is important to remark that, if in the considered clique there is at least another node $i_{\tilde k}$ forming with $i_k$ an edge included in some other clique, let us say a $(g^\prime,r)$-clique, $q_{i_k,g^\prime}^{(r)}$ cannot be linearized, since the product corresponding to that $(g^\prime,r)$-clique would be made of state probabilities conditioned to the state of both $i_k$ and $i_{\tilde k}$ (not of $i_k$ only, as in Eq.~(\ref{eq:q_lin})). Indeed, those states in which $i_{\tilde k}$ is in state $I$, would give ${\cal O}(1)$ terms, for both numerator and denominator would be ${\cal O}(\epsilon)$. This is the reason why in markovian models, when aiming to account for the dynamical correlations within some subsets of nodes (e.g., cliques), a consistent expression for the critical point can be given only when those subsets are edge-disjoint.

Returning to the derivation, since $P_{i_1\dots i_n,g}^{S\dots S}$ is fixed by the normalization condition,
\begin{equation}
  P_{i_1\dots i_n,g}^{S\dots S}=1-\sum_{\scriptstyle {\begin{array}{c}{\{\sigma_1, \dots, \sigma_n\}}\\ \neq\{S,\dots, S\}\end{array}}}P_{i_1\dots i_n,g}^{\sigma_{i_1} \dots \sigma_{i_n}}
\end{equation}

\noindent we only need the linearized equations for arrival states with at least one infected node, i.e., $\{\sigma_{i_1}^\prime,\dots ,\sigma_{i_n}^\prime\}\neq\{S,\dots ,S\}$. Retaining the $\cal{O}(\epsilon)$ terms in $\phi_{i_k,g}$, after some algebra, we find
\begin{widetext}
\begin{align}
  \notag P_{i_1\dots i_n,g}^{\sigma_{i_1}^\prime \dots \sigma_{i_n}^\prime} &= \sum_{\scriptstyle {\begin{array}{c}{\{\sigma_{i_1}, \dots, \sigma_{i_n}\}}\\ \neq\{S,\dots, S\}\end{array}}} P_{i_1\dots i_n,g}^{\sigma_{i_1} \dots \sigma_{i_n}} \left[\left(1-\mu\right)^{N_{I\rightarrow I}}~\mu^{N_{I\rightarrow S}}\left(1-w_{N_I,g}^{(n-1)}\right)^{N_{S\rightarrow I}}\left(w_{N_I,g}^{(n-1)}\right)^{N_{S\rightarrow S}}\right]
  \\
  \notag &+~{\mathbb{1}}[\exists! k:\sigma_{i_k}^\prime=I]\left\{\sum_{(g^\prime,r)\neq (g,n-1)}\sum_{\scriptstyle {\begin{array}{c}{\{j_1,\dots,j_r\}}\\ \in\Gamma_{i_k,g^\prime}^{(r)}\end{array}}}\left[\sum_{l=1}^r \frac{1-w_{l,g^\prime}^{(r)}}{l!\left(r-l\right)!}\sum_{k_1\neq\dots\neq k_r=1}^r P_{j_{k_1}\dots j_{k_l}j_{k_{l+1}}\dots j_{k_r}i_k,g^\prime}^{I\dots IS\dots SS}\right]\right.
  \\
  &\hspace{66pt}\left. +\sum_{\scriptstyle {\begin{array}{c}{\{j_1,\dots,j_{n-1}\}}\\ \in\Gamma_{i_k,g}^{(n-1)}\backslash \{\neg i_k\}\end{array}}} \left[\sum_{l=1}^{n-1} \frac{1-w_{l,g}^{(n-1)}}{l!\left(n-1-l\right)!}\sum_{k_1\neq\dots\neq k_{n-1}=1}^{n-1} P_{j_{k_1}\dots j_{k_l}j_{k_{l+1}}\dots j_{k_{n-1}}i_k,g}^{I\dots IS\dots SS}\right]\right\}
  + \mathcal{O}(\epsilon^2)
  \label{eq:DE_n_clique_lin}
\end{align}
\end{widetext}

\noindent where $N_{\sigma\rightarrow \sigma^\prime}=\left\vert\{i_{k=1,\dots ,n}\vert\sigma_{i_k}=\sigma, \sigma_{i_k}^\prime=\sigma^\prime\}\right\vert$ is the number of nodes going from state $\sigma$ to state $\sigma^\prime$. The terms in curly brackets derive from those transitions starting in state $\{S,\dots ,S\}$ and arriving to a state with exactly one infected node.

In particular, Eq.~(\ref{eq:P_i}), the dynamic equation for a single node, becomes
\begin{align}
  \notag P_i^I = P_i^I(1-&\mu)+\sum_{(g,r)} \sum_{\scriptstyle {\begin{array}{c}{\{j_1,\dots,j_r\}}\\ \in\Gamma_{i,g}^{(r)}\end{array}}}\left[\sum_{l=1}^r \frac{1-w_{l,g}^{(r)}}{l!\left(r-l\right)!}~\right.
  \\
  &\left.\times\sum_{k_1\neq\dots\neq k_r=1}^r P_{j_{k_1}\dots j_{k_l}j_{k_{l+1}}\dots j_{k_r}i,g}^{I\dots IS\dots SS}\right]
  + \mathcal{O}(\epsilon^2)
  \label{eq:P_i_lin}
\end{align}

At this point, to get an expression for the critical threshold, we put Eq.~(\ref{eq:P_i_lin}) in the form of an eigenvalue equation for the vector of single-node probabilities $\mathbf{P}^I=\left(\{P_i^I\}_{i\in V}\right)$. To this purpose, given a $(\cdot ,n)$-clique $\{i_1,\dots ,i_n\}$, we need to express every joint probability over its nodes states as a linear combination of the marginal probabilities,  $P_{i_1}^I,\dots,P_{i_n}^I$. Eq.~(\ref{eq:DE_n_clique_lin}) provides a linearized equation for each of the $2^n-1$ unknown state probabilities, hence a system admitting a unique solution. At this point, instead of the $n$~equations for the transition to a state with a single node in state~$I$ (second term in Eq.~(\ref{eq:DE_n_clique_lin})), we use the $n$ consistency relations for the marginal probabilities, i.e., $P_{i_k}^I=\sum_{\{\sigma_{\neg i_k}\}} P_{i_k\{\neg i_k\},g}^{I\{\sigma_{\neg i_k}\}}$, $\forall k\in\{1,\dots,n\}$. In this way, the system is still made of $2^n-1$ equations, hence is determined, but now it includes the marginal probabilities. Eventually, with some algebra, one gets the decomposition with its linear coefficients. Alternatively, asserted the uniqueness of the solution of the linear system, the problem can be approached in the other way around. That is, firstly expressing each of the joint probabilities as the most general linear combination of the marginal probabilities and then inserting them into the $2^n-1$ equations. Doing so, we get a new, determined linear system whose unknowns are the linear coefficients of the original system. In the end, given a $(g,n)$-clique $\{i_1,\dots,i_n\}$, the most general and proper linear decomposition of $P_{i_1\dots i_n,g}^{\sigma_{i_1} \dots \sigma_{i_n}}$ in terms of $P_{i_1}^I,\dots,P_{i_n}^I$, reads
\begin{equation}
  P_{i_1\dots i_n,g}^{\sigma_{i_1} \dots \sigma_{i_n}} = X_{N_I,g}^{(n-1)}\sum_{k:\sigma_{i_k}=I}P_{i_k}^I + Y_{N_I,g}^{(n-1)}\sum_{k:\sigma_{i_k}=S}P_{i_k}^I
  + \mathcal{O}(\epsilon^2)
  \label{eq:P_decom}
\end{equation}

\noindent where we can take $Y_{n,g}^{(n-1)}=0$, being null the term it multiplies, as there are no nodes in state $S$ for $N_I = n$. We get one coefficient for $N_I=n$ and two coefficients for each $N_I\in\{1,\dots ,n-1\}$, leading to $2n-1$ of them in total. Summing for every order $n$ from $2$ to $m_0$ for $g=0$, and from $3$ to $m_1$ for $g=1$, we get a maximum of ${m_0}^2+{m_1}^2-5$ linear coefficients to fix. These coefficients, as functions of all the $m_1$~microscopic parameters of the model, $\{\beta^{(s)}\}_{s=1,\dots ,{m_1-1}}$ and $\mu$, weigh the probability of finding a clique in a given state, when the system approaches the critical point.

Once all the coefficients have been found by insertion of Eq.~(\ref{eq:P_decom}) in the original linear system of $2^n-1$ equations, we substitute them in Eq.~(\ref{eq:P_i_lin}) to finally get an eigenvalue equation. To this end, we define the set of ${\mathcal D({\mathcal K})}$-dependent adjacency matrices $\left\{\left\{A^{(0,r)}\right\}_{r\in\{1,\dots ,m_0-1\}},\left\{A^{(1,r)}\right\}_{r\in\{2,\dots ,m_1-1\}}\right\}$, such that $A_{ij}^{(g,r)}$ equals $1$ if nodes $i$ and $j$ share a common incident $(g,r+1)$-clique in ${\mathcal D({\mathcal K})}$, and $0$ otherwise. Besides, we define the $(g,r)$ degree of node $i$, $k_i^{(g,r)}$, as the number of $(g,r+1)$-cliques incident on node $i$ in ${\mathcal D({\mathcal K})}$, computed as $k_i^{(g,r)}=\sum_{j=1}^N A_{ij}^{(g,r)}$. Being the decomposition edge-disjoint, only one of those matrices can have a non-zero element in the position corresponding to a given pair of nodes. Consequently, the ${\mathcal D({\mathcal K})}$-independent adjacency matrix of ${\mathcal K}^{(1)}$, the underlying graph of ${\mathcal K}$, is simply obtained by the sum of all those ${\mathcal D({\mathcal K})}$-dependent adjacency matrices. It also follows that the degree $k_i$ of node $i$ in ${\mathcal K}^{(1)}$ is computed as $k_i=k_i^{(0)}+k_i^{(1)}$, where $k_i^{(0)}=\sum_{r=1}^{m_0}rk_i^{(0,r)}$ and $k_i^{(1)}=\sum_{r=2}^{m_1}rk_i^{(1,r)}$ are the total number of neighbors of $i$ within, respectively, $(0,\cdot)$-cliques and $(1,\cdot)$-cliques.

Substituting Eq.~(\ref{eq:P_decom}) in Eq.~(\ref{eq:P_i_lin}), and doing some algebra and combinatorics, we get
\begin{equation}
  M\mathbf{P}^I = D\mathbf{P}^I
  \label{eq:eigen}
\end{equation}

\noindent where we have defined the matrices $M$ and $D$, of elements
\begin{align}
  M_{ij} &= \sum_{(g,r)}A_{ij}^{(g,r)}\sum_{l=1}^r\left(1-w_{l,g}^{(r)}\right)\left[\binom{r-1}{l-1}X_{l,g}^{(r)}+\binom{r-1}{r-l-1}Y_{l,g}^{(r)}\right]
  \label{eq:M} \\
  D_{ij} &= \delta_{ij}\left[\mu-\sum_{(g,r)} k_i^{(g,r)}\sum_{l=1}^r \binom{r}{l}\left(1-w_{l,g}^{(r)}\right)Y_{l,g}^{(r)}\right]
  \label{eq:D}
\end{align}

\noindent being $\delta_{ij}$ the elements of the $N \times N$ identity matrix. Equations~(\ref{eq:eigen}) to~(\ref{eq:D}) define a generalized eigenvalue problem. The form taken by $M$ and $D$ is easily understood in this way. Each sum multiplying $A_{ij}^{(g,r)}$ in $M_{ij}$ represents the marginal contribution to the infection of node $i$ coming from its neighbor $j$ through the $(g,r+1)$-clique they share. Given $j$ in state $I$, $\binom{r-1}{a}$ is the number of ways in which $a$ out of $r-1$ nodes can be chosen to be in state $I$ ($a=l-1$) or $S$ ($a=r-l-1$). Similarly, each sum multiplying $k_i^{(g,r)}$ in $D_{ii}$ represents the contribution coming from all the configurations of any of the $(g,r+1)$-cliques incident on node $i$, in which $i$ is in state~$S$. $\binom{r}{l}$ is the number of ways in which $l$ out of $r$ nodes can be chosen to be in state $I$.

Now, $M$ is non-negative. Indeed, for any fixed $(g,r)$, the sum multiplying $A_{ij}^{(g,r)}$ must be positive whenever $A_{ij}^{(g,r)}>0$, since it represents the by-definition positive contribution to the infection of a node ($i$) coming from one of its neighbors ($j$). Moreover, being $\mathcal{K}$ undirected and so $M$ symmetric, it follows that $M$ is also irreducible \cite{meyer2000matrix}. Looking now at $D$, which is diagonal, the non-negativity of $M$ implies the diagonal elements of $D$ to be positive. Indeed, Eq.~(\ref{eq:eigen}) holds for any value of the microscopic parameters; so let us suppose $\mu=0$. Since the non-zero elements of $M$ are positive $\forall\mu\in\left[0,1\right]$, the sum multiplying $k_i^{(g,r)}$ in $D_{ij}$ must be negative whenever $A_{ij}^{(g,r)}>0$, proving any diagonal element of $D$ to be positive. Therefore, $D$ is invertible and its inverse as well, with elements $\left[D^{-1}\right]_{ii}=\left(D_{ii}\right)^{-1}$, $\forall i=1,\dots ,N$. Applying $D^{-1}$ to both sides of Eq.~(\ref{eq:eigen}), we finally get the sought eigenvalue equation,
\begin{equation}
  M^\prime\mathbf{P}^I = \mathbf{P}^I
  \label{eq:eigen_2}
\end{equation}

\noindent where we have defined the matrix $M^\prime\equiv D^{-1} M$, of elements $M^\prime_{ij}=M_{ij}/D_{ii}$. Thus, $M^\prime$ is a non-negative irreducible matrix as well and, by the Perron-Frobenius theorem \cite{meyer2000matrix}, it admits a unique leading eigenvector $\mathbf{P}^I_{\star}$. This is the only one associated with the largest eigenvalue $\Lambda_{\textup{max}}\left(M^\prime\right)$ and the only one with all its entries positive, hence representing the unique physically acceptable expected state of the system at the onset of the epidemic. Fixed the values of the recovery probability, $\mu$, and the higher-order infection probabilities, $\{\beta^{(s)}\}_{s>1}$, the critical threshold $\beta^{(1)}_{\textup{cr}}$ is implicitly found as the smallest non-negative value of $\beta^{(1)}$ such that
\begin{equation}
  \Lambda_{\textup{max}}\left(M^\prime\right) = 1
  \label{eq:epi_th}
\end{equation}

\vspace{4.ex}
\noindent{\fontfamily{cmss}\selectfont{\textbf{Mean-field approximation.}}} The homogeneous mean-field (MF) approximation of Eq.~(\ref{eq:P_i}) is found by neglecting both the state correlations and the local structural heterogeneity among the nodes, i.e., regarding every node as the ``average node'' in the structure \cite{gomez2011nonperturbative}. Given any $(g,r)$-clique $\{j_1,\dots ,j_r\}$, it follows $P_{j_1\dots j_l j_{l+1}\dots j_r,g}^{I\dots IS\dots S}=\rho^{l}\left(1-\rho\right)^{r-l}$; and, for any node $i$, $k_i^{(g,r)}=\bar{k}^{(g,r)}$, where $\bar{k}^{(g,r)}$ is the average value of the $(g,r)$-degree of the nodes in the structure. Thus, Eq.~(\ref{eq:P_i}) becomes
\begin{equation}
  \rho\left(t+1\right) = \rho\left(t\right)\left(1-\mu\right)+\left(1-\rho\left(t\right)\right)\left(1-\prod_{(g,r)}\bar{q}_{g}^{(r)}\right)
  \label{eq:rho_mf}
\end{equation}

\noindent where
\begin{equation}
  \bar{q}_{g}^{(r)} = \left[1-\sum_{l=1}^r\binom{r}{l}\left(1-w_{l,g}^{(r)}\right)\rho^{l}\left(1-\rho\right)^{r-l}\right]^{{\bar{k}^{(g,r)}}}
  \label{eq:q_mf}
\end{equation}

The stationary solution is then got imposing $\rho\left(t+1\right)=\rho\left(t\right)$ in Eq.~(\ref{eq:rho_mf}). The linearization around the epidemic-free state is then implemented by taking $\rho = \epsilon \ll 1$. Looking at Eq.~(\ref{eq:q_mf}), the only ${\cal O}(\epsilon)$ terms are given by $l=1$, whatever the couple $(g,r)$. That is, only the pair-wise probability $\beta^{(1)}$ contributes in the MF approximation. With few algebra, one gets the renowned formula
\begin{equation}
  \beta^{(1)}_{\textup{cr}} = \frac{\mu}{\bar{k}}
  \label{eq:epi_th_mf}
\end{equation}

\noindent where $\bar{k}=\frac{1}{N}\sum_{i=1}^N k_i$, is the average degree of a node in ${\mathcal K}^{(1)}$.

More generally, Eq.~(\ref{eq:epi_th_mf}) holds for any model treating the nodes states as independent, thus including the MMCA approximation of the MECLE, the MMCA and MF approximations of the simplicial ELE, and also, expect for substituting $\bar{k}$ with $\xbar{k^2}/\bar{k}$, the heterogeneous MF approximation \cite{gomez2011nonperturbative} of both. The same result is found in continuous time \cite{landry2020effect}.

\vspace{4.ex}
\noindent{\fontfamily{cmss}\selectfont{\textbf{Numerical simulations.}}} The equilibrium value of the prevalence $\rho$ is computed using synchronous Monte Carlo simulations and the quasistationary state (QS) method \cite{ferreira2012epidemic}. In the specific case of the simplicial $2$-complexes, for each node $i\in V$, a simulated time step proceeds as follows: (1) if $i$ is currently infected, it recovers with probability $\mu$; (2) if $i$ is currently susceptible, (2.1) it gets infected with probability $1-(1-\beta^{(1)})^{n_i^{(1)}}$, being $n_i^{(1)}$ the number of currently infected neighbors of $i$ through edges; (2.2) if $i$ is still susceptible after sub-step (2.1), it gets infected with probability $1-(1-\beta^{(2)})^{n_i^{(2)}}$, being $n_i^{(2)}$ the number of currently infected couples of neighbors of $i$ through triangles. For higher dimensions, $n>2$, step (2) consists of $n-2$ additional sub-steps, analogously defined and ordered as shown here.

In accordance with the QS method, every time the absorbing state $\rho=0$ is reached, it is replaced by one of the previously stored active states of the system, i.e., one of those states with at least an active individual. Since for finite systems, when approaching the critical point, a large number of realizations end up in the absorbing state, the QS method properly reduces to a single run the wasteful method of performing many simulations. We have made use of $50$~stored active states and an update probability of $0.25$. We have given the systems a transient time of $10^5$~time steps, and then calculated $\rho$ as an average over $2\times 10^4$~additional time steps.

\vspace{4.ex}
\noindent{\fontfamily{cmss}\selectfont{\textbf{Connectivity structures.}}} We have referred to various interaction structures. For the numerical evaluation, we have made use of synthetic simplicial $2$-complexes with around $N=10^4$ nodes, presenting dissimilar structural properties: regular structures, as the clique complexes built from a triangular lattice; homogeneous structures, generated from the random simplicial complex model \cite{iacopini2019simplicial}, in which both the $(\cdot,1)$- and the $(\cdot,2)$-degree follow a Poisson distribution; heterogeneous structures, derived from the Dorogovtsev-Mendes model \cite{dorogovtsev2001size}, having $(\cdot,1)$- and $(\cdot,2)$-degree distributions nearly following power-laws with exponent around $3$.

\section*{\large{Data availability}}
Data sharing not applicable to this article as no datasets were generated or analyzed during the current study.

\section*{\large{Code availability}}
The code for estimating a minimal edge-disjoint edge clique cover (EECC) of a graph has been implemented as the DisjointCliqueCover.jl\cite{burgio2021eecc} package for the Julia language, available at \href{https://github.com/giubuig/DisjointCliqueCover.jl}{github} and archived at \href{https://doi.org/10.5281/zenodo.4723748}{zenodo.org}.

\endgroup


\begin{thebibliography}{10}
\expandafter\ifx\csname url\endcsname\relax
  \def\url#1{\texttt{#1}}\fi
\expandafter\ifx\csname urlprefix\endcsname\relax\def\urlprefix{URL }\fi
\providecommand{\bibinfo}[2]{#2}
\providecommand{\eprint}[2][]{\url{#2}}

\bibitem{anderson1992infectious}
\bibinfo{author}{Anderson, R.~M.} \& \bibinfo{author}{May, R.~M.}
\newblock \emph{\bibinfo{title}{Infectious diseases of humans: dynamics and
  control}} (\bibinfo{publisher}{Oxford university press},
  \bibinfo{year}{1992}).

\bibitem{daley1965stochastic}
\bibinfo{author}{Daley, D.~J.} \& \bibinfo{author}{Kendall, D.~G.}
\newblock \bibinfo{title}{Stochastic rumours}.
\newblock \emph{\bibinfo{journal}{IMA Journal of Applied Mathematics}}
  \textbf{\bibinfo{volume}{1}}, \bibinfo{pages}{42--55} (\bibinfo{year}{1965}).

\bibitem{rogers2010diffusion}
\bibinfo{author}{Rogers, E.~M.}
\newblock \emph{\bibinfo{title}{Diffusion of innovations}}
  (\bibinfo{publisher}{Simon and Schuster}, \bibinfo{year}{2010}).

\bibitem{katz1966personal}
\bibinfo{author}{Katz, E.} \& \bibinfo{author}{Lazarsfeld, P.~F.}
\newblock \emph{\bibinfo{title}{Personal Influence, The part played by people
  in the flow of mass communications}} (\bibinfo{publisher}{Transaction
  Publishers}, \bibinfo{year}{1966}).

\bibitem{valente1995network}
\bibinfo{author}{Valente, T.~W.}
\newblock \emph{\bibinfo{title}{Network models of the diffusion of
  innovations}}.
\newblock \bibinfo{number}{303.484 V3} (\bibinfo{publisher}{Hampton Press},
  \bibinfo{year}{1995}).

\bibitem{cowan2004network}
\bibinfo{author}{Cowan, R.} \& \bibinfo{author}{Jonard, N.}
\newblock \bibinfo{title}{Network structure and the diffusion of knowledge}.
\newblock \emph{\bibinfo{journal}{Journal of Economic Dynamics and Control}}
  \textbf{\bibinfo{volume}{28}}, \bibinfo{pages}{1557--1575}
  (\bibinfo{year}{2004}).

\bibitem{pastor2001epidemic}
\bibinfo{author}{Pastor-Satorras, R.} \& \bibinfo{author}{Vespignani, A.}
\newblock \bibinfo{title}{Epidemic spreading in scale-free networks}.
\newblock \emph{\bibinfo{journal}{Physical Review Letters}}
  \textbf{\bibinfo{volume}{86}}, \bibinfo{pages}{3200} (\bibinfo{year}{2001}).

\bibitem{watts2007influentials}
\bibinfo{author}{Watts, D.~J.} \& \bibinfo{author}{Dodds, P.~S.}
\newblock \bibinfo{title}{Influentials, networks, and public opinion
  formation}.
\newblock \emph{\bibinfo{journal}{Journal of Consumer Research}}
  \textbf{\bibinfo{volume}{34}}, \bibinfo{pages}{441--458}
  (\bibinfo{year}{2007}).

\bibitem{watts1998collective}
\bibinfo{author}{Watts, D.~J.} \& \bibinfo{author}{Strogatz, S.~H.}
\newblock \bibinfo{title}{Collective dynamics of ‘small-world’networks}.
\newblock \emph{\bibinfo{journal}{Nature}} \textbf{\bibinfo{volume}{393}},
  \bibinfo{pages}{440--442} (\bibinfo{year}{1998}).

\bibitem{bianconi2014triadic}
\bibinfo{author}{Bianconi, G.}, \bibinfo{author}{Darst, R.~K.},
  \bibinfo{author}{Iacovacci, J.} \& \bibinfo{author}{Fortunato, S.}
\newblock \bibinfo{title}{Triadic closure as a basic generating mechanism of
  communities in complex networks}.
\newblock \emph{\bibinfo{journal}{Physical Review E}}
  \textbf{\bibinfo{volume}{90}}, \bibinfo{pages}{042806}
  (\bibinfo{year}{2014}).

\bibitem{palla2005uncovering}
\bibinfo{author}{Palla, G.}, \bibinfo{author}{Der{\'e}nyi, I.},
  \bibinfo{author}{Farkas, I.} \& \bibinfo{author}{Vicsek, T.}
\newblock \bibinfo{title}{Uncovering the overlapping community structure of
  complex networks in nature and society}.
\newblock \emph{\bibinfo{journal}{Nature}} \textbf{\bibinfo{volume}{435}},
  \bibinfo{pages}{814--818} (\bibinfo{year}{2005}).

\bibitem{battiston2020networks}
\bibinfo{author}{Battiston, F.} \emph{et~al.}
\newblock \bibinfo{title}{Networks beyond pairwise interactions: structure and
  dynamics}.
\newblock \emph{\bibinfo{journal}{Physics Reports}}  (\bibinfo{year}{2020}).

\bibitem{burgio2020evolution}
\bibinfo{author}{Burgio, G.}, \bibinfo{author}{Matamalas, J.~T.},
  \bibinfo{author}{G{\'o}mez, S.} \& \bibinfo{author}{Arenas, A.}
\newblock \bibinfo{title}{Evolution of cooperation in the presence of
  higher-order interactions: from networks to hypergraphs}.
\newblock \emph{\bibinfo{journal}{Entropy}} \textbf{\bibinfo{volume}{22}},
  \bibinfo{pages}{744} (\bibinfo{year}{2020}).

\bibitem{dai2020d}
\bibinfo{author}{Dai, X.} \emph{et~al.}
\newblock \bibinfo{title}{D-dimensional oscillators in simplicial structures:
  odd and even dimensions display different synchronization scenarios}.
\newblock \emph{\bibinfo{journal}{arXiv preprint arXiv:2010.14976}}
  (\bibinfo{year}{2020}).

\bibitem{ghorbanchian2020higher}
\bibinfo{author}{Ghorbanchian, R.}, \bibinfo{author}{Restrepo, J.~G.},
  \bibinfo{author}{Torres, J.~J.} \& \bibinfo{author}{Bianconi, G.}
\newblock \bibinfo{title}{Higher-order simplicial synchronization of coupled
  topological signals}.
\newblock \emph{\bibinfo{journal}{arXiv preprint arXiv:2011.00897}}
  (\bibinfo{year}{2020}).

\bibitem{tadic2021hidden}
\bibinfo{author}{Tadi{\'c}, B.} \& \bibinfo{author}{Gupte, N.}
\newblock \bibinfo{title}{Hidden geometry and dynamics of complex networks:
  Spin reversal in nanoassemblies with pairwise and triangle-based interactions
  (a)}.
\newblock \emph{\bibinfo{journal}{EPL (Europhysics Letters)}}
  \textbf{\bibinfo{volume}{132}}, \bibinfo{pages}{60008}
  (\bibinfo{year}{2021}).

\bibitem{sun2020renormalization}
\bibinfo{author}{Sun, H.}, \bibinfo{author}{Ziff, R.~M.} \&
  \bibinfo{author}{Bianconi, G.}
\newblock \bibinfo{title}{Renormalization group theory of percolation on
  pseudofractal simplicial and cell complexes}.
\newblock \emph{\bibinfo{journal}{Physical Review E}}
  \textbf{\bibinfo{volume}{102}}, \bibinfo{pages}{012308}
  (\bibinfo{year}{2020}).

\bibitem{andjelkovic2020topology}
\bibinfo{author}{Andjelkovi{\'c}, M.}, \bibinfo{author}{Tadi{\'c}, B.} \&
  \bibinfo{author}{Melnik, R.}
\newblock \bibinfo{title}{The topology of higher-order complexes associated
  with brain hubs in human connectomes}.
\newblock \emph{\bibinfo{journal}{Scientific Reports}}
  \textbf{\bibinfo{volume}{10}}, \bibinfo{pages}{1--10} (\bibinfo{year}{2020}).

\bibitem{bretto2013hypergraph}
\bibinfo{author}{Bretto, A.}
\newblock \emph{\bibinfo{title}{Hypergraph theory. An introduction.}}
  (\bibinfo{publisher}{Springer}, \bibinfo{year}{2013}).

\bibitem{lambiotte2019networks}
\bibinfo{author}{Lambiotte, R.}, \bibinfo{author}{Rosvall, M.} \&
  \bibinfo{author}{Scholtes, I.}
\newblock \bibinfo{title}{From networks to optimal higher-order models of
  complex systems}.
\newblock \emph{\bibinfo{journal}{Nature Physics}}
  \textbf{\bibinfo{volume}{15}}, \bibinfo{pages}{313--320}
  (\bibinfo{year}{2019}).

\bibitem{jonsson2007simplicial}
\bibinfo{author}{Jonsson, J.}
\newblock \emph{\bibinfo{title}{Simplicial complexes of graphs}}
  (\bibinfo{publisher}{Springer Science \& Business Media},
  \bibinfo{year}{2007}).

\bibitem{granovetter1978threshold}
\bibinfo{author}{Granovetter, M.}
\newblock \bibinfo{title}{Threshold models of collective behavior}.
\newblock \emph{\bibinfo{journal}{American Journal of Sociology}}
  \textbf{\bibinfo{volume}{83}}, \bibinfo{pages}{1420--1443}
  (\bibinfo{year}{1978}).

\bibitem{centola2007complex}
\bibinfo{author}{Centola, D.} \& \bibinfo{author}{Macy, M.}
\newblock \bibinfo{title}{Complex contagions and the weakness of long ties}.
\newblock \emph{\bibinfo{journal}{American Journal of Sociology}}
  \textbf{\bibinfo{volume}{113}}, \bibinfo{pages}{702--734}
  (\bibinfo{year}{2007}).

\bibitem{centola2010spread}
\bibinfo{author}{Centola, D.}
\newblock \bibinfo{title}{The spread of behavior in an online social network
  experiment}.
\newblock \emph{\bibinfo{journal}{Science}} \textbf{\bibinfo{volume}{329}},
  \bibinfo{pages}{1194--1197} (\bibinfo{year}{2010}).

\bibitem{melnik2013multi}
\bibinfo{author}{Melnik, S.}, \bibinfo{author}{Ward, J.~A.},
  \bibinfo{author}{Gleeson, J.~P.} \& \bibinfo{author}{Porter, M.~A.}
\newblock \bibinfo{title}{Multi-stage complex contagions}.
\newblock \emph{\bibinfo{journal}{Chaos: An Interdisciplinary Journal of
  Nonlinear Science}} \textbf{\bibinfo{volume}{23}}, \bibinfo{pages}{013124}
  (\bibinfo{year}{2013}).

\bibitem{watts2002simple}
\bibinfo{author}{Watts, D.~J.}
\newblock \bibinfo{title}{A simple model of global cascades on random
  networks}.
\newblock \emph{\bibinfo{journal}{Proceedings of the National Academy of
  Sciences USA}} \textbf{\bibinfo{volume}{99}}, \bibinfo{pages}{5766--5771}
  (\bibinfo{year}{2002}).

\bibitem{guilbeault2018complex}
\bibinfo{author}{Guilbeault, D.}, \bibinfo{author}{Becker, J.} \&
  \bibinfo{author}{Centola, D.}
\newblock \bibinfo{title}{Complex contagions: A decade in review}.
\newblock In \emph{\bibinfo{booktitle}{Complex spreading phenomena in social
  systems}}, \bibinfo{pages}{3--25} (\bibinfo{publisher}{Springer},
  \bibinfo{year}{2018}).

\bibitem{ugander2012structural}
\bibinfo{author}{Ugander, J.}, \bibinfo{author}{Backstrom, L.},
  \bibinfo{author}{Marlow, C.} \& \bibinfo{author}{Kleinberg, J.}
\newblock \bibinfo{title}{Structural diversity in social contagion}.
\newblock \emph{\bibinfo{journal}{Proceedings of the National Academy of
  Sciences USA}} \textbf{\bibinfo{volume}{109}}, \bibinfo{pages}{5962--5966}
  (\bibinfo{year}{2012}).

\bibitem{weng2012competition}
\bibinfo{author}{Weng, L.}, \bibinfo{author}{Flammini, A.},
  \bibinfo{author}{Vespignani, A.} \& \bibinfo{author}{Menczer, F.}
\newblock \bibinfo{title}{Competition among memes in a world with limited
  attention}.
\newblock \emph{\bibinfo{journal}{Scientific Reports}}
  \textbf{\bibinfo{volume}{2}}, \bibinfo{pages}{335} (\bibinfo{year}{2012}).

\bibitem{iacopini2019simplicial}
\bibinfo{author}{Iacopini, I.}, \bibinfo{author}{Petri, G.},
  \bibinfo{author}{Barrat, A.} \& \bibinfo{author}{Latora, V.}
\newblock \bibinfo{title}{Simplicial models of social contagion}.
\newblock \emph{\bibinfo{journal}{Nature Communications}}
  \textbf{\bibinfo{volume}{10}}, \bibinfo{pages}{1--9} (\bibinfo{year}{2019}).

\bibitem{landry2020effect}
\bibinfo{author}{Landry, N.~W.} \& \bibinfo{author}{Restrepo, J.~G.}
\newblock \bibinfo{title}{The effect of heterogeneity on hypergraph contagion
  models}.
\newblock \emph{\bibinfo{journal}{Chaos: An Interdisciplinary Journal of
  Nonlinear Science}} \textbf{\bibinfo{volume}{30}}, \bibinfo{pages}{103117}
  (\bibinfo{year}{2020}).

\bibitem{st2021bursty}
\bibinfo{author}{St-Onge, G.}, \bibinfo{author}{Sun, H.},
  \bibinfo{author}{Allard, A.}, \bibinfo{author}{H{\'e}bert-Dufresne, L.} \&
  \bibinfo{author}{Bianconi, G.}
\newblock \bibinfo{title}{Bursty exposure on higher-order networks leads to
  nonlinear infection kernels}.
\newblock \emph{\bibinfo{journal}{arXiv preprint arXiv:2101.07229}}
  (\bibinfo{year}{2021}).

\bibitem{bodo2016sis}
\bibinfo{author}{Bod{\'o}, {\'A}.}, \bibinfo{author}{Katona, G.~Y.} \&
  \bibinfo{author}{Simon, P.~L.}
\newblock \bibinfo{title}{Sis epidemic propagation on hypergraphs}.
\newblock \emph{\bibinfo{journal}{Bulletin of Mathematical Biology}}
  \textbf{\bibinfo{volume}{78}}, \bibinfo{pages}{713--735}
  (\bibinfo{year}{2016}).

\bibitem{jhun2019simplicial}
\bibinfo{author}{Jhun, B.}, \bibinfo{author}{Jo, M.} \& \bibinfo{author}{Kahng,
  B.}
\newblock \bibinfo{title}{Simplicial sis model in scale-free uniform
  hypergraph}.
\newblock \emph{\bibinfo{journal}{Journal of Statistical Mechanics: Theory and
  Experiment}} \textbf{\bibinfo{volume}{2019}}, \bibinfo{pages}{123207}
  (\bibinfo{year}{2019}).

\bibitem{de2020social}
\bibinfo{author}{de~Arruda, G.~F.}, \bibinfo{author}{Petri, G.} \&
  \bibinfo{author}{Moreno, Y.}
\newblock \bibinfo{title}{Social contagion models on hypergraphs}.
\newblock \emph{\bibinfo{journal}{Physical Review Research}}
  \textbf{\bibinfo{volume}{2}}, \bibinfo{pages}{023032} (\bibinfo{year}{2020}).

\bibitem{matamalas2020abrupt}
\bibinfo{author}{Matamalas, J.~T.}, \bibinfo{author}{G{\'o}mez, S.} \&
  \bibinfo{author}{Arenas, A.}
\newblock \bibinfo{title}{Abrupt phase transition of epidemic spreading in
  simplicial complexes}.
\newblock \emph{\bibinfo{journal}{Physical Review Research}}
  \textbf{\bibinfo{volume}{2}}, \bibinfo{pages}{012049} (\bibinfo{year}{2020}).

\bibitem{henkel2008non}
\bibinfo{author}{Henkel, M.}, \bibinfo{author}{Hinrichsen, H.} \&
  \bibinfo{author}{L{\"u}beck, S.}
\newblock \emph{\bibinfo{title}{Non-equilibrium phase transitions: volume 1:
  absorbing phase transitions}} (\bibinfo{publisher}{Springer Science \&
  Business Media}, \bibinfo{year}{2008}).

\bibitem{matamalas2018effective}
\bibinfo{author}{Matamalas, J.~T.}, \bibinfo{author}{Arenas, A.} \&
  \bibinfo{author}{G{\'o}mez, S.}
\newblock \bibinfo{title}{Effective approach to epidemic containment using link
  equations in complex networks}.
\newblock \emph{\bibinfo{journal}{Science Advances}}
  \textbf{\bibinfo{volume}{4}}, \bibinfo{pages}{eaau4212}
  (\bibinfo{year}{2018}).

\bibitem{gomez2010discrete}
\bibinfo{author}{G{\'o}mez, S.}, \bibinfo{author}{Arenas, A.},
  \bibinfo{author}{Borge-Holthoefer, J.}, \bibinfo{author}{Meloni, S.} \&
  \bibinfo{author}{Moreno, Y.}
\newblock \bibinfo{title}{Discrete-time markov chain approach to contact-based
  disease spreading in complex networks}.
\newblock \emph{\bibinfo{journal}{EPL (Europhysics Letters)}}
  \textbf{\bibinfo{volume}{89}}, \bibinfo{pages}{38009} (\bibinfo{year}{2010}).

\bibitem{dorogovtsev2001size}
\bibinfo{author}{Dorogovtsev, S.~N.}, \bibinfo{author}{Mendes, J.~F.} \&
  \bibinfo{author}{Samukhin, A.~N.}
\newblock \bibinfo{title}{Size-dependent degree distribution of a scale-free
  growing network}.
\newblock \emph{\bibinfo{journal}{Physical Review E}}
  \textbf{\bibinfo{volume}{63}}, \bibinfo{pages}{062101}
  (\bibinfo{year}{2001}).

\bibitem{barrat2021social}
\bibinfo{author}{Barrat, A.}, \bibinfo{author}{de~Arruda, G.~F.},
  \bibinfo{author}{Iacopini, I.} \& \bibinfo{author}{Moreno, Y.}
\newblock \bibinfo{title}{Social contagion on higher-order structures}.
\newblock \emph{\bibinfo{journal}{arXiv preprint arXiv:2103.03709}}
  (\bibinfo{year}{2021}).

\bibitem{dankulov2019spectral}
\bibinfo{author}{Dankulov, M.~M.}, \bibinfo{author}{Tadi{\'c}, B.} \&
  \bibinfo{author}{Melnik, R.}
\newblock \bibinfo{title}{Spectral properties of hyperbolic nanonetworks with
  tunable aggregation of simplexes}.
\newblock \emph{\bibinfo{journal}{Physical Review E}}
  \textbf{\bibinfo{volume}{100}}, \bibinfo{pages}{012309}
  (\bibinfo{year}{2019}).

\bibitem{meyer2000matrix}
\bibinfo{author}{Meyer, C.~D.}
\newblock \emph{\bibinfo{title}{Matrix analysis and applied linear algebra}},
  vol.~\bibinfo{volume}{71} (\bibinfo{publisher}{SIAM}, \bibinfo{year}{2000}).

\bibitem{gomez2011nonperturbative}
\bibinfo{author}{G{\'o}mez, S.}, \bibinfo{author}{G{\'o}mez-Gardenes, J.},
  \bibinfo{author}{Moreno, Y.} \& \bibinfo{author}{Arenas, A.}
\newblock \bibinfo{title}{Nonperturbative heterogeneous mean-field approach to
  epidemic spreading in complex networks}.
\newblock \emph{\bibinfo{journal}{Physical Review E}}
  \textbf{\bibinfo{volume}{84}}, \bibinfo{pages}{036105}
  (\bibinfo{year}{2011}).

\bibitem{ferreira2012epidemic}
\bibinfo{author}{Ferreira, S.~C.}, \bibinfo{author}{Castellano, C.} \&
  \bibinfo{author}{Pastor-Satorras, R.}
\newblock \bibinfo{title}{Epidemic thresholds of the
  susceptible-infected-susceptible model on networks: A comparison of numerical
  and theoretical results}.
\newblock \emph{\bibinfo{journal}{Physical Review E}}
  \textbf{\bibinfo{volume}{86}}, \bibinfo{pages}{041125}
  (\bibinfo{year}{2012}).

\end{thebibliography}

\begingroup
\fontsize{8pt}{9pt}\selectfont

\section*{\large{Acknowledgments}}
G.B. acknowledges financial support from the European Union's Horizon 2020 research and innovation program under the Marie Sk\l{}odowska-Curie Grant Agreement No. 945413. We acknowledge support by Ministerio de Econom\'{\i}a y Competitividad (PGC2018-094754-BC21, FIS2017-90782-REDT and RED2018-102518-T), Generalitat de Catalunya (2017SGR-896 and 2020PANDE00098), and Universitat Rovira i Virgili (2019PFR-URV-B2-41). A.A.\ acknowledges also ICREA Academia and the James S.\ McDonnell Foundation (220020325).

\section*{\large{Author contributions}}
GB, AA, SG and JTM designed the research and wrote the manuscript. GB and JTM made the analytical calculations. GB performed the numerical analysis and simulations.

\section*{\large{Competing interests}}
The authors declare no competing interests.

\endgroup

\onecolumngrid

\renewcommand{\theequation}{S\arabic{equation}}
\setcounter{equation}{0}

\renewcommand{\figurename}{{Supplementary Figure}}
\newcounter{SMfigure}
\renewcommand{\thefigure}{\arabic{SMfigure}}

\vspace{2cm}
\section*{Supplementary Note 1.\quad Robustness of MECLE under changes in the EECC}
\label{sec:robustness}

Here we assess the dependence of the predictions made by the MECLE on the use of different EECCs found for a given SC ${\cal K}$. If the maximal cliques of the underlying graph ${\cal K}^{(1)}$ of the latter are all mutually edge-disjoint, then there is a unique EECC. Otherwise, two cases must be distinguished: (i) ${\cal K}$ is $0$-connected, so the only non-edge-disjoint cliques in ${\cal K}^{(1)}$ are cliques also in ${\cal K}$ (i.e., they are $(0,\cdot)$-cliques); (ii) ${\cal K}$ is non-$0$-connected, so the non-edge-disjoint cliques in ${\cal K}^{(1)}$ can be either cliques or simplices in ${\cal K}$ (i.e., $(0,\cdot)$- or $(1,\cdot)$-cliques, respectively). The MECLE strictly applies only to the case (i) but, as shown in the main text, it keeps working well when the $0$-connectedness is only slightly broken. In order to prove how robust is the MECLE in both cases, we consider structures possessing high proportions of edges shared by multiple cliques. Indeed, the higher is the number of times edges are shared, the higher is the probability that the proposed heuristic returns (proportionally) different EECCs. We recall that the algorithm is not deterministic only when, at some point, among the cliques with the minimum score, there are multiple ones having maximum order, among which one is randomly chosen (step 4(b) of the proposed heuristic).

Let us first focus on case (i). The best way to prove the robustness of the MECLE is, in this case, considering a graph (i.e., a simplicial $2$-complex), so that the number of overlapping maximal $(0,\cdot)$-cliques is maximized; otherwise, some of them would be $(1,\cdot)$-cliques, which in this case are supposed $0$-connected and hence not affecting the edge covering. In Supplementary Fig.~\ref{fig:robustness_0con} we show and discuss the results obtained computing several EECCs of a graph generated by the Dorogovtsev-Mendes model \cite{dorogovtsev2001size}. Such a graph has no tree-like portions. Despite its high rate of edge overlap, the prediction made by the MECLE is substantially independent from the used covering.

Going to case (ii), we consider several EECCs of the clique complex of the Dorogovtsev-Mendes network used in (i). The clique complex is obtained by converting all the $3$-cliques of the network in $2$-faces, so it is strongly non-$0$-connected. Note that the MECLE could be unreliable for such structure, since the algorithm destroys so many $2$-faces that the resulting EECCs are too far from the original structure. The results in Supplementary Fig.~\ref{fig:robustness_non0con} show the low variability among the EECCs provided by our heuristic and, consequently, among the predictions made by the MECLE, even for highly non-$0$-connected structures. Accordingly, such variability becomes negligible when the SC only slightly breaks the $0$-connectedness condition, so that is sufficient to consider just one EECC. This is the case for the SCs considered in Fig.~5, for which the MECLE is still reliable.

\newpage

\section*{Supplementary Note 2.\quad MECLE for simplicial 2-complexes}
\label{sec:mecle_sim2com}

We here illustrate the form taken by the MECLE model when the interaction structure is a simplicial $2$-complex, i.e., the $(3,3)$ implementation of the model. Cliques and faces can only have order $n=2,3$.

The evolution of the probability $P_i^I$ for node~$i$ being infected is governed by Eq.~(8), which now takes the form
\begin{equation}
  P_i^I\left(t+1\right) = P_i^I\left(t\right)\left(1-\mu\right)+P_i^S\left(t\right)\left(1-q_{i,0}^{(1)}q_{i,0}^{(2)}q_{i,1}^{(2)}\right)
  \label{eq:P_i_sim2com}
\end{equation}

\noindent where
\begin{subequations}
\begin{align}
  q_{i,0}^{(1)} = &\prod_{\scriptstyle {j\in\Gamma_{i,0}^{(1)}}} \left[1-\beta^{(1)}P_{j\vert i,0}^{I\vert S}\right]
  \\
  q_{i,g}^{(2)} = &\prod_{\scriptstyle {\{j, k\}\in\Gamma_{i,g}^{(2)}}} \left[1-\beta^{(1)}\left(P_{jk\vert i,g}^{IS\vert S}+P_{jk\vert i,g}^{SI\vert S}\right)-\left(1-\left(1-\beta^{(1)}\right)^2\left(1-g\beta^{(2)}\right)\right)P_{jk\vert i,g}^{II\vert S}\right]
\end{align}
\label{eq:q_sim2com}
\end{subequations}

\indent According to Eq.~(2), the state of a $(0,1)$-clique $\{i,j\}$ is governed by the following equations
\begin{subequations}
\begin{align}
  \notag P_{ij,0}^{II}(t+1) &= P_{ij,0}^{SS}(t)\left(1-q_{i(j),0}^{(1)}q_{i,0}^{(2)}q_{i,1}^{(2)}\right)\left(1-q_{j(i),0}^{(1)}q_{j,0}^{(2)}q_{j,1}^{(2)}\right)
  \\
  \notag &+ P_{ij,0}^{SI}(t)\left(1-(1-\beta^{(1)})q_{i(j),0}^{(1)}q_{i,0}^{(2)}q_{i,1}^{(2)}\right)\left(1-\mu\right)
  \\
  \notag &+ P_{ij,0}^{IS}(t)\left(1-\mu\right)\left(1-(1-\beta^{(1)})q_{j(i),0}^{(1)}q_{j,0}^{(2)}q_{j,1}^{(2)}\right)
  \\
  &+ P_{ij,0}^{II}(t)\left(1-\mu\right)^2
  \\
  \notag P_{ij,0}^{IS}(t+1) &= P_{ij,0}^{SS}(t)\left(1-q_{i(j),0}^{(1)}q_{i,0}^{(2)}q_{i,1}^{(2)}\right)\left(q_{j(i),0}^{(1)}q_{j,0}^{(2)}q_{j,1}^{(2)}\right)
  \\
  \notag &+ P_{ij,0}^{SI}(t)\left(1-(1-\beta^{(1)})q_{i(j),0}^{(1)}q_{i,0}^{(2)}q_{i,1}^{(2)}\right)\mu
  \\
  \notag &+ P_{ij,0}^{IS}(t)\left(1-\mu\right)\left((1-\beta^{(1)})q_{j(i),0}^{(1)}q_{j,0}^{(2)}q_{j,1}^{(2)}\right)
  \\
  &+ P_{ij,0}^{II}(t)\left(1-\mu\right)\mu
\end{align}
\label{eq:edge_sim2com}
\end{subequations}

\noindent where $q_{i(j),0}^{(1)}$ coincides with $q_{i,0}^{(1)}$ except for excluding the $(0,1)$-clique $\{i,j\}$ from the product, and analogously for the other similar terms.

The state of a $(0,2)$-clique $\{i,j,k\}$ follows the equations
\begin{subequations}
\begin{align}
  \notag P_{ijk,0}^{III}(t+1) &= P_{ijk,0}^{SSS}(t)\left(1-q_{i,0}^{(1)}q_{i\left(jk\right),0}^{(2)}q_{i,1}^{(2)}\right)\left(1-q_{j,0}^{(1)}q_{j\left(ik\right),0}^{(2)}q_{j,1}^{(2)}\right)\left(1-q_{k,0}^{(1)}q_{k\left(ij\right),0}^{(2)}q_{k,1}^{(2)}\right)
  \\
  \notag &+ P_{ijk,0}^{SSI}(t)\left(1-\left(1-\beta^{(1)}\right)q_{i,0}^{(1)}q_{i\left(jk\right),0}^{(2)}q_{i,1}^{(2)}\right)\left(1-\left(1-\beta^{(1)}\right)q_{j,0}^{(1)}q_{j\left(ik\right),0}^{(2)}q_{j,1}^{(2)}\right)\left(1-\mu\right)
  \\
  \notag &+ P_{ijk,0}^{SIS}(t)\left(1-\left(1-\beta^{(1)}\right)q_{i,0}^{(1)}q_{i\left(jk\right),0}^{(2)}q_{i,1}^{(2)}\right)\left(1-\mu\right)\left(1-\left(1-\beta^{(1)}\right)q_{k,0}^{(1)}q_{k\left(ij\right),0}^{(2)}q_{k,1}^{(2)}\right)
  \\
  \notag &+ P_{ijk,0}^{ISS}(t)\left(1-\mu\right)\left(1-\left(1-\beta^{(1)}\right)q_{j,0}^{(1)}q_{j\left(ik\right),0}^{(2)}q_{j,1}^{(2)}\right)\left(1-\left(1-\beta^{(1)}\right)q_{k,0}^{(1)}q_{k\left(ij\right),0}^{(2)}q_{k,1}^{(2)}\right)
  \\
  \notag &+ P_{ijk,0}^{SII}(t)\left(1-\left(1-\beta^{(1)}\right)^2 q_{i,0}^{(1)}q_{i\left(jk\right),0}^{(2)}q_{i,1}^{(2)}\right)\left(1-\mu\right)^2
  \\
  \notag &+ P_{ijk,0}^{ISI}(t)\left(1-\mu\right)\left(1-\left(1-\beta^{(1)}\right)^2 q_{j,0}^{(1)}q_{j\left(ik\right),0}^{(2)}q_{j,1}^{(2)}\right)\left(1-\mu\right)
  \\
  \notag &+ P_{ijk,0}^{IIS}(t)\left(1-\mu\right)^2\left(1-\left(1-\beta^{(1)}\right)^2 q_{k,0}^{(1)}q_{k\left(ij\right),0}^{(2)}q_{k,1}^{(2)}\right)
  \\
  &+ P_{ijk,0}^{III}(t)\left(1-\mu\right)^3
  \\
  \notag P_{ijk,0}^{IIS}(t+1) &= P_{ijk,0}^{SSS}(t)\left(1-q_{i,0}^{(1)}q_{i\left(jk\right),0}^{(2)}q_{i,1}^{(2)}\right)\left(1-q_{j,0}^{(1)}q_{j\left(ik\right),0}^{(2)}q_{j,1}^{(2)}\right)\left(q_{k,0}^{(1)}q_{k\left(ij\right),0}^{(2)}q_{k,1}^{(2)}\right)
  \\
  \notag &+ P_{ijk,0}^{SSI}(t)\left(1-\left(1-\beta^{(1)}\right)q_{i,0}^{(1)}q_{i\left(jk\right),0}^{(2)}q_{i,1}^{(2)}\right)\left(1-\left(1-\beta^{(1)}\right)q_{j,0}^{(1)}q_{j\left(ik\right),0}^{(2)}q_{j,1}^{(2)}\right)\mu
  \\
  \notag &+ P_{ijk,0}^{SIS}(t)\left(1-\left(1-\beta^{(1)}\right)q_{i,0}^{(1)}q_{i\left(jk\right),0}^{(2)}q_{i,1}^{(2)}\right)\left(1-\mu\right)\left(\left(1-\beta^{(1)}\right)q_{k,0}^{(1)}q_{k\left(ij\right),0}^{(2)}q_{k,1}^{(2)}\right)
  \\
  \notag &+ P_{ijk,0}^{ISS}(t)\left(1-\mu\right)\left(1-\left(1-\beta^{(1)}\right)q_{j,0}^{(1)}q_{j\left(ik\right),0}^{(2)}q_{j,1}^{(2)}\right)\left(\left(1-\beta^{(1)}\right)q_{k,0}^{(1)}q_{k\left(ij\right),0}^{(2)}q_{k,1}^{(2)}\right)
  \\
  \notag &+ P_{ijk,0}^{SII}(t)\left(1-\left(1-\beta^{(1)}\right)^2 q_{i,0}^{(1)}q_{i\left(jk\right),0}^{(2)}q_{i,1}^{(2)}\right)\left(1-\mu\right)\mu
  \\
  \notag &+ P_{ijk,0}^{ISI}(t)\left(1-\mu\right)\left(1-\left(1-\beta^{(1)}\right)^2 q_{j,0}^{(1)}q_{j\left(ik\right),0}^{(2)}q_{j,1}^{(2)}\right)\mu
  \\
  \notag &+ P_{ijk,0}^{IIS}(t)\left(1-\mu\right)^2\left(\left(1-\beta^{(1)}\right)^2 q_{k,0}^{(1)}q_{k\left(ij\right),0}^{(2)}q_{k,1}^{(2)}\right)
  \\
  &+ P_{ijk,0}^{III}(t)\left(1-\mu\right)^2 \mu
  \\
  \notag P_{ijk,0}^{ISS}(t+1) &= P_{ijk,0}^{SSS}(t)\left(1-q_{i,0}^{(1)}q_{i\left(jk\right),0}^{(2)}q_{i,1}^{(2)}\right)\left(q_{j,0}^{(1)}q_{j\left(ik\right),0}^{(2)}q_{j,1}^{(2)}\right)\left(q_{k,0}^{(1)}q_{k\left(ij\right),0}^{(2)}q_{k,1}^{(2)}\right)
  \\
  \notag &+ P_{ijk,0}^{SSI}(t)\left(1-\left(1-\beta^{(1)}\right)q_{i,0}^{(1)}q_{i\left(jk\right),0}^{(2)}q_{i,1}^{(2)}\right)\left(\left(1-\beta^{(1)}\right)q_{j,0}^{(1)}q_{j\left(ik\right),0}^{(2)}q_{j,1}^{(2)}\right)\mu
  \\
  \notag &+ P_{ijk,0}^{SIS}(t)\left(1-\left(1-\beta^{(1)}\right)q_{i,0}^{(1)}q_{i\left(jk\right),0}^{(2)}q_{i,1}^{(2)}\right)\mu\left(\left(1-\beta^{(1)}\right)q_{k,0}^{(1)}q_{k\left(ij\right),0}^{(2)}q_{k,1}^{(2)}\right)
  \\
  \notag &+ P_{ijk,0}^{ISS}(t)\left(1-\mu\right)\left(\left(1-\beta^{(1)}\right)q_{j,0}^{(1)}q_{j\left(ik\right),0}^{(2)}q_{j,1}^{(2)}\right)\left(\left(1-\beta^{(1)}\right)q_{k,0}^{(1)}q_{k\left(ij\right),0}^{(2)}q_{k,1}^{(2)}\right)
  \\
  \notag &+ P_{ijk,0}^{SII}(t)\left(1-\left(1-\beta^{(1)}\right)^2 q_{i,0}^{(1)}q_{i\left(jk\right),0}^{(2)}q_{i,1}^{(2)}\right)\mu^2
  \\
  \notag &+ P_{ijk,0}^{ISI}(t)\left(1-\mu\right)\left(\left(1-\beta^{(1)}\right)^2 q_{j,0}^{(1)}q_{j\left(ik\right),0}^{(2)}q_{j,1}^{(2)}\right)\mu
  \\
  \notag &+ P_{ijk,0}^{IIS}(t)\left(1-\mu\right)\mu\left(\left(1-\beta^{(1)}\right)^2 q_{k,0}^{(1)}q_{k\left(ij\right),0}^{(2)}q_{k,1}^{(2)}\right)
  \\
  &+ P_{ijk,0}^{III}(t)\left(1-\mu\right)\mu^2
\end{align}
\end{subequations}
\label{eq:triangle0_sim2com}

\noindent where $q_{i(jk),0}^{(2)}$ coincides with $q_{i,0}^{(2)}$ except for excluding the $(0,2)$-clique $\{i,j,k\}$ from the product, and analogously for the other similar terms.

Finally, for a $(1,2)$-clique $\{i,j,k\}$, we get the following equations
\begin{subequations}
\begin{align}
  \notag P_{ijk,1}^{III}(t+1) &= P_{ijk,1}^{SSS}(t)\left(1-q_{i,0}^{(1)}q_{i,0}^{(2)}q_{i\left(jk\right),1}^{(2)}\right)\left(1-q_{j,0}^{(1)}q_{j,0}^{(2)}q_{j\left(ik\right),1}^{(2)}\right)\left(1-q_{k,0}^{(1)}q_{k,0}^{(2)}q_{k\left(ij\right),1}^{(2)}\right)
  \\
  \notag &+ P_{ijk,1}^{SSI}(t)\left(1-\left(1-\beta^{(1)}\right)q_{i,0}^{(1)}q_{i,0}^{(2)}q_{i\left(jk\right),1}^{(2)}\right)\left(1-\left(1-\beta^{(1)}\right)q_{j,0}^{(1)}q_{j,0}^{(2)}q_{j\left(ik\right),1}^{(2)}\right)\left(1-\mu\right)
  \\
  \notag &+ P_{ijk,1}^{SIS}(t)\left(1-\left(1-\beta^{(1)}\right)q_{i,0}^{(1)}q_{i,0}^{(2)}q_{i\left(jk\right),1}^{(2)}\right)\left(1-\mu\right)\left(1-\left(1-\beta^{(1)}\right)q_{k,0}^{(1)}q_{k,0}^{(2)}q_{k\left(ij\right),1}^{(2)}\right)
  \\
  \notag &+ P_{ijk,1}^{ISS}(t)\left(1-\mu\right)\left(1-\left(1-\beta^{(1)}\right)q_{j,0}^{(1)}q_{j,0}^{(2)}q_{j\left(ik\right),1}^{(2)}\right)\left(1-\left(1-\beta^{(1)}\right)q_{k,0}^{(1)}q_{k,0}^{(2)}q_{k\left(ij\right),1}^{(2)}\right)
  \\
  \notag &+ P_{ijk,1}^{SII}(t)\left(1-\left(1-\beta^{(1)}\right)^2\left(1-\beta^{(2)}\right)q_{i,0}^{(1)}q_{i,0}^{(2)}q_{i\left(jk\right),1}^{(2)}\right)\left(1-\mu\right)^2
  \\
  \notag &+ P_{ijk,1}^{ISI}(t)\left(1-\mu\right)\left(1-\left(1-\beta^{(1)}\right)^2\left(1-\beta^{(2)}\right)q_{j,0}^{(1)}q_{j,0}^{(2)}q_{j\left(ik\right),1}^{(2)}\right)\left(1-\mu\right)
  \\
  \notag &+ P_{ijk,1}^{IIS}(t)\left(1-\mu\right)^2\left(1-\left(1-\beta^{(1)}\right)^2\left(1-\beta^{(2)}\right)q_{k,0}^{(1)}q_{k,0}^{(2)}q_{k\left(ij\right),1}^{(2)}\right)
  \\
  &+ P_{ijk,1}^{III}(t)\left(1-\mu\right)^3
  \\
  \notag P_{ijk,1}^{IIS}(t+1) &= P_{ijk,1}^{SSS}(t)\left(1-q_{i,0}^{(1)}q_{i,0}^{(2)}q_{i\left(jk\right),1}^{(2)}\right)\left(1-q_{j,0}^{(1)}q_{j,0}^{(2)}q_{j\left(ik\right),1}^{(2)}\right)\left(q_{k,0}^{(1)}q_{k,0}^{(2)}q_{k\left(ij\right),1}^{(2)}\right)
  \\
  \notag &+ P_{ijk,1}^{SSI}(t)\left(1-\left(1-\beta^{(1)}\right)q_{i,0}^{(1)}q_{i,0}^{(2)}q_{i\left(jk\right),1}^{(2)}\right)\left(1-\left(1-\beta^{(1)}\right)q_{j,0}^{(1)}q_{j,0}^{(2)}q_{j\left(ik\right),1}^{(2)}\right)\mu
  \\
  \notag &+ P_{ijk,1}^{SIS}(t)\left(1-\left(1-\beta^{(1)}\right)q_{i,0}^{(1)}q_{i,0}^{(2)}q_{i\left(jk\right),1}^{(2)}\right)\left(1-\mu\right)\left(\left(1-\beta^{(1)}\right)q_{k,0}^{(1)}q_{k,0}^{(2)}q_{k\left(ij\right),1}^{(2)}\right)
  \\
  \notag &+ P_{ijk,1}^{ISS}(t)\left(1-\mu\right)\left(1-\left(1-\beta^{(1)}\right)q_{j,0}^{(1)}q_{j,0}^{(2)}q_{j\left(ik\right),1}^{(2)}\right)\left(\left(1-\beta^{(1)}\right)q_{k,0}^{(1)}q_{k,0}^{(2)}q_{k\left(ij\right),1}^{(2)}\right)
  \\
  \notag &+ P_{ijk,1}^{SII}(t)\left(1-\left(1-\beta^{(1)}\right)^2\left(1-\beta^{(2)}\right)q_{i,0}^{(1)}q_{i,0}^{(2)}q_{i\left(jk\right),1}^{(2)}\right)\left(1-\mu\right)\mu
  \\
  \notag &+ P_{ijk,1}^{ISI}(t)\left(1-\mu\right)\left(1-\left(1-\beta^{(1)}\right)^2\left(1-\beta^{(2)}\right)q_{j,0}^{(1)}q_{j,0}^{(2)}q_{j\left(ik\right),1}^{(2)}\right)\mu
  \\
  \notag &+ P_{ijk,1}^{IIS}(t)\left(1-\mu\right)^2\left(\left(1-\beta^{(1)}\right)^2\left(1-\beta^{(2)}\right)q_{k,0}^{(1)}q_{k,0}^{(2)}q_{k\left(ij\right),1}^{(2)}\right)
  \\
  &+ P_{ijk,1}^{III}(t)\left(1-\mu\right)^2 \mu
  \\
  \notag P_{ijk,1}^{ISS}(t+1) &= P_{ijk,1}^{SSS}(t)\left(1-q_{i,0}^{(1)}q_{i,0}^{(2)}q_{i\left(jk\right),1}^{(2)}\right)\left(q_{j,0}^{(1)}q_{j,0}^{(2)}q_{j\left(ik\right),1}^{(2)}\right)\left(q_{k,0}^{(1)}q_{k,0}^{(2)}q_{k\left(ij\right),1}^{(2)}\right)
  \\
  \notag &+ P_{ijk,1}^{SSI}(t)\left(1-\left(1-\beta^{(1)}\right)q_{i,0}^{(1)}q_{i,0}^{(2)}q_{i\left(jk\right),1}^{(2)}\right)\left(\left(1-\beta^{(1)}\right)q_{j,0}^{(1)}q_{j,0}^{(2)}q_{j\left(ik\right),1}^{(2)}\right)\mu
  \\
  \notag &+ P_{ijk,1}^{SIS}(t)\left(1-\left(1-\beta^{(1)}\right)q_{i,0}^{(1)}q_{i,0}^{(2)}q_{i\left(jk\right),1}^{(2)}\right)\mu\left(\left(1-\beta^{(1)}\right)q_{k,0}^{(1)}q_{k,0}^{(2)}q_{k\left(ij\right),1}^{(2)}\right)
  \\
  \notag &+ P_{ijk,1}^{ISS}(t)\left(1-\mu\right)\left(\left(1-\beta^{(1)}\right)q_{j,0}^{(1)}q_{j,0}^{(2)}q_{j\left(ik\right),1}^{(2)}\right)\left(\left(1-\beta^{(1)}\right)q_{k,0}^{(1)}q_{k,0}^{(2)}q_{k\left(ij\right),1}^{(2)}\right)
  \\
  \notag &+ P_{ijk,1}^{SII}(t)\left(1-\left(1-\beta^{(1)}\right)^2\left(1-\beta^{(2)}\right)q_{i,0}^{(1)}q_{i,0}^{(2)}q_{i\left(jk\right),1}^{(2)}\right)\mu^2
  \\
  \notag &+ P_{ijk,1}^{ISI}(t)\left(1-\mu\right)\left(\left(1-\beta^{(1)}\right)^2\left(1-\beta^{(2)}\right)q_{j,0}^{(1)}q_{j,0}^{(2)}q_{j\left(ik\right),1}^{(2)}\right)\mu
  \\
  \notag &+ P_{ijk,1}^{IIS}(t)\left(1-\mu\right)\mu\left(\left(1-\beta^{(1)}\right)^2\left(1-\beta^{(2)}\right)q_{k,0}^{(1)}q_{k,0}^{(2)}q_{k\left(ij\right),1}^{(2)}\right)
  \\
  &+ P_{ijk,1}^{III}(t)\left(1-\mu\right)\mu^2
\end{align}
\label{eq:triangle1_sim2com}
\end{subequations}

\noindent where $q_{i(jk),1}^{(2)}$ coincides with $q_{i,1}^{(2)}$ except for excluding the $(1,2)$-clique $\{i,j,k\}$ from the product, and analogously for the other similar terms.

\newpage

\section*{Supplementary Note 3.\quad Epidemic threshold for highly symmetrical structures}
\label{sec:epi_th_sym}

Given the high accuracy of the MECLE ---as shown in the main text---, here we make further predictions for some particularly symmetrical structures, for which a closed form of $\beta^{(1)}_{\textup{cr}}$, expressing it in terms of the higher-order infection probabilities, $\{\beta^{(s)}\}_{s>1}$, and the recovery probability, $\mu$, can be provided from Eq.~(19). As a particular case, we show the monotonous decrease of $\beta^{(1)}_{\textup{cr}}$ with respect to $\beta^{(2)}$ in the herein examined $2$-dimensional SCs. We stress that such dependence, as like as any other one, of $\beta^{(1)}_{\textup{cr}}$ on the higher-order couplings, is completely overlooked by models treating the nodes' states as uncorrelated.

\begin{paragraph}{Regular SCs.}
For a SC such that $k_i^{(g,r)}=k^{(g,r)}$, $\forall i\in V$, $\forall (g,r)$, since all the non-zero elements of $M^\prime$ are equal to the same constant, from the Collatz–Wielandt formula \cite{meyer2000matrix} it follows
\begin{equation}
  \Lambda_{\textup{max}}\left(M^\prime\right)=\sum_j M^\prime_{ij} = \frac{\sum_{(g,r)}rk^{(g,r)}\sum_{l=1}^r\left(1-w_{l,g}^{(r)}\right)\left[\binom{r-1}{l-1}X_{l,g}^{(r)}+\binom{r-1}{r-l-1}Y_{l,g}^{(r)}\right]}{\mu-\sum_{(g,r)} k^{(g,r)}\sum_{l=1}^r \binom{r}{l}\left(1-w_{l,g}^{(r)}\right)Y_{l,g}^{(r)}}
  \label{eq:epi_th_regular}
\end{equation}

\noindent where both equalities hold for any chosen $i\in V$. Using Eq.~(19), $\Lambda_{\textup{max}}\left(M^\prime\right)=1$, one can solve with respect to $\beta^{(1)}$ and express $\beta^{(1)}_{\textup{cr}}$ in terms of all the other parameters. The decreasing of $\beta^{(1)}_{\textup{cr}}$ with $\beta^{(2)}$ is shown in Supplementary Fig.~\ref{fig:regular_diagram} for two classes of regular simplicial $2$-complexes. In particular, the periodic triangular clique complex, considered in the main text, falls within this class of structures. Note there is no dependence on the number of nodes, $N=\vert V\vert$.
\end{paragraph}

\begin{paragraph}{Friendship SCs.} As an opposite case, we consider now extremely heterogeneous SCs. A Friendship graph $\textup{F}_n$ is a Windmill graph $\textup{Wd}(m,n)$ whose ``sails'' are cliques of order $m=3$. It consists of $N=2n+1$ nodes, where $n$ is the number of $3$-cliques incident on the central node. Starting from $\textup{F}_n$, we convert a fraction $p_\triangle$ of the $n$ $3$-cliques in $2$-faces. The central node has $k^{(0,2)}=(1-p_\triangle)n\equiv n_0$ and $k^{(1,2)}=p_\triangle n\equiv n_1$. Then, $2 n_0$ of the peripheral nodes have each $k^{(0,2)}=1$ and $k^{(1,2)}=0$, while the remaining $2 n_1$ have $k^{(1,2)}=1$ and $k^{(0,2)}=0$. The greater $N$ (hence, $n$), the higher the heterogeneity between the central node and the peripheral ones. In the large $N$ limit, the average number of neighbors $\bar k=3\frac{N-1}{N}$ tends to $3$, whereas any higher $m$-moment diverges as $N^{m-1}$. Accordingly, we expect the epidemic threshold to vanish in that limit.

In order to find a closed expression for the epidemic threshold $\beta^{(1)}_{\textup{cr}}$, we take advantage of the nearly block structure featured by matrix $M^\prime$. It can be partitioned as
\begin{equation}
  M^\prime=
  \begin{pmatrix}
    B & P \\
    C & 0
  \end{pmatrix}
\end{equation}

\noindent where $B$ is a $(N-1)\times (N-1)$ block diagonal matrix, where each $2\times 2$ block corresponds to the peripheral edge of a $(0,3)$- or a $(1,3)$-clique. That is, $B$ can be put in the form
\begin{equation}
  B=\textup{diag}(\underbrace{B_0,\dots,B_0}_{n_0},\underbrace{B_1,\dots,B_1}_{n_1})
\end{equation}

\noindent where
\begin{equation}
  B_g=
  \begin{pmatrix}
  0 & M^\prime_{(P,g)} \\
  M^\prime_{(P,g)} & 0
  \end{pmatrix}
\end{equation}

\noindent and
\begin{equation}
  M^\prime_{(P,g)}=~\frac{\beta^{(1)}\left(X_{1,g}^{(2)}+Y_{1,g}^{(2)}\right)+\left(1-\left(1-\beta^{(1)}\right)^2\left(1-g\beta^{(2)}\right)\right)X_{2,g}^{(2)}}{\mu-2\beta^{(1)} Y_{1,g}^{(2)}-\left(1-\left(1-\beta^{(1)}\right)^2\left(1-g\beta^{(2)}\right)\right)Y_{2,g}^{(2)}}
\end{equation}

\noindent for $g=0,1$. $P$ is a $(N-1)\times 1$ matrix whose elements equal $M^\prime_{(P,0)}$ or $M^\prime_{(P,1)}$ depending on whether the peripheral node corresponding to the considered row participates, respectively, to a $(0,3)$- or a $(1,3)$-clique. Similarly, $C$ is a $1\times (N-1)$ matrix whose elements equal $M^\prime_{(C,0)}$ or $M^\prime_{(C,1)}$ depending on whether the central node participates, respectively, to a $(0,3)$- or a $(1,3)$-clique with the peripheral node corresponding to the considered column; where
\begin{equation}
  M^\prime_{(C,g)} =~ \frac{\beta^{(1)}\left(X_{1,g}^{(2)}+Y_{1,g}^{(2)}\right)+\left(1-\left(1-\beta^{(1)}\right)^2\left(1-g\beta^{(2)}\right)\right)X_{2,g}^{(2)}}{\mu-\sum_{g=0,1} n_g\left[2\beta^{(1)} Y_{1,g}^{(2)}+\left(1-\left(1-\beta^{(1)}\right)^2\left(1-g\beta^{(2)}\right)\right)Y_{2,g}^{(2)}\right]}
\end{equation}

\noindent for $g=0,1$.

We now compute the determinant of \begin{equation}
  M^\prime-\lambda I_N =
  \begin{pmatrix}
  B-\lambda I_{N-1} & P
  \\
  C & -\lambda \end{pmatrix}
\end{equation}

\noindent where $I_N$ is the $N\times N$ identity matrix. Thanks to the Schur complement formula \cite{meyer2000matrix}, we can compute it as
\begin{equation}
  \textup{det}(M^\prime-\lambda I_{N}) =
  \textup{det}(B-\lambda I_{N-1}) \left[-\lambda-C\left(B-\lambda I_{N-1}\right)^{-1} P\right]
\label{det_windmill}
\end{equation}

\noindent Using the properties of block diagonal matrices \cite{meyer2000matrix},
\begin{equation}
  \textup{det}(B-\lambda I_{N-1})=\left(\lambda^2-{M^\prime_{(P,0)}}^2\right)^{n_0} \left(\lambda^2-{M^\prime_{(P,1)}}^2\right)^{n_1}
\end{equation}

\noindent implying that $\lambda_{(P,g)} \equiv M^\prime_{(P,g)}$, $\forall g\in\{0,1\}$, solves $\textup{det}(M^\prime-\lambda I_{N})=0$; and therefore one between $\lambda_{(P,0)}$ and $\lambda_{(P,1)}$ is the leading eigenvalue of $B$. However, receiving contributions from peripheral nodes only, the latter can be shown to never coincide with the largest eigenvalue of $M^\prime$. In particular, when both $n_0>1$ and $n_1>1$, we already know this is true, for the largest one is a simple eigenvalue \cite{meyer2000matrix}. Therefore, let us suppose $\lambda\neq\lambda_{(P,g)}$, $\forall g\in\{0,1\}$, and look for $\Lambda_{\textup{max}}\left(M^\prime\right)$ in the other factor, the one containing the contribution coming also from the central node. It is easily found that
\begin{equation}
  \left(B-\lambda I_{N-1}\right)^{-1} = \textup{diag}(\underbrace{\tilde{B_0}^{-1},\dots,\tilde{B_0}^{-1}}_{n_0},\underbrace{\tilde{B_1}^{-1},\dots,\tilde{B_1}^{-1}}_{n_1})
\end{equation}

\noindent being
\begin{equation}
  \tilde{B_g}^{-1} =
  \frac{1}{\lambda^2-{M^\prime_{(P,g)}}^2}
  \begin{pmatrix}
    -\lambda&-M^\prime_{(P,g)}
    \\
    -M^\prime_{(P,g)}&-\lambda
  \end{pmatrix}
\end{equation}

\noindent the inverse matrix of $\tilde{B_g}=B_g-\lambda I_{2}$, $g=0,1$. With a few algebra, it follows
\begin{equation}
  C \left(B-\lambda I_{N-1}\right)^{-1} P = -2\left(n_0\frac{M^\prime_{(P,0)}M^\prime_{(C,0)}}{\lambda-M^\prime_{(P,0)}}+n_1\frac{M^\prime_{(P,1)}M^\prime_{(C,1)}}{\lambda-M^\prime_{(P,1)}}\right)
\end{equation}

We now impose $\textup{det}(M^\prime-\lambda I_{N})=0$ which, using the previous result $\lambda\neq\lambda_{(P,g)}$, $\forall g\in\{0,1\}$, reduces to $-\lambda-C \left(B-\lambda I_{N-1}\right)^{-1} P=0$. This can be rearranged in the form
\begin{align}
  \notag\lambda^3-\lambda^2\left(M^\prime_{(P,0)}+M^\prime_{(P,1)}\right)+\lambda\left(M^\prime_{(P,0)}M^\prime_{(P,1)}-2n_0 M^\prime_{(P,0)}M^\prime_{(C,0)}-2n_1 M^\prime_{(P,1)}M^\prime_{(C,1)}\right)&
  \\
  +~2M^\prime_{(P,0)}M^\prime_{(P,1)}\left(n_0M^\prime_{(C,0)}+n_1 M^\prime_{(C,1)}\right)=0&
  \label{eq:eigenvalue_windmill}
\end{align}

\noindent We finally look for $\lambda=\Lambda_{\textup{max}}\left(M^\prime\right)$ among the solutions of Eq.~(\ref{eq:eigenvalue_windmill}). As before, we solve $\Lambda_{\textup{max}}\left(M^\prime\right)=1$ with respect to $\beta^{(1)}$ to find $\beta^{(1)}_{\textup{cr}}$ as a function of the microscopic parameters, $\beta^{(2)}$ and $\mu$, and of $N$ and $p_\triangle$. Results are shown in Supplementary Fig.~\ref{fig:windmill_diagram}, where several values of $N$ are explored for $p_\triangle=0.5$. There is shown that the epidemic threshold vanishes in the limit of large~$N$, while always decreasing with $\beta^{(2)}$. Interestingly, the dependence on $\beta^{(2)}$ grows with $N$, hence with the degree disparity between the central node and the peripheral ones. This, together with the weaker dependence found for regular structures (Supplementary Fig.~\ref{fig:regular_diagram}), suggests that the dependence of $\beta^{(1)}_{\textup{cr}}$ on $\beta^{(2)}$ grows with the heterogeneity of connections. In fact, as shown in Fig.~4, a very similar dependence is found for Dorogovtsev-Mendes SCs. This indicates that, despite their simplicity, the Friendship SCs are able to capture some important dynamical properties of more complex heterogeneous structures. To notice that, a strong correlation between edge-degree, $k^{(0,1)}$, and triangle-degree, $k^{(1,2)}$, of a node exists in both Dorogovtsev-Mendes SCs and Friendship SCs.
\end{paragraph}
\vspace{2em}

\newpage

\section*{Supplementary Note 4.\quad Continuous-time limit of MECLE for simplicial 2-complexes}
\label{sec:mecle_sim2com_cont}

It is possible to derive the continuous-time equations of the SIS dynamic in SCs as a limit of the MECLE model. Here, we show this process for the particular case of interaction structures arranged in simplicial $2$-complexes, i.e., the continuous-time version of Eqs.~\ref{eq:P_i_sim2com}-\ref{eq:triangle1_sim2com}. In order to take the continuous-time limit, we make the substitutions
\begin{subequations}
\begin{align}
  \notag \mu &\longrightarrow \mu\,\Delta t
  \\
  \notag \beta^{(s)}& \longrightarrow \beta^{(s)}\,\Delta t
\end{align}
\end{subequations}
where now $\mu$ and $\beta^{(s)}$ represent rates, i.e., probabilities per unit time, instead of the original discrete-time probabilities \cite{gomez2011nonperturbative}. Then, we take the limit $\Delta t\rightarrow 0$, which means neglecting all those terms in Eqs.~\ref{eq:P_i_sim2com}-\ref{eq:triangle1_sim2com} appearing as second or greater powers of $\Delta t$ or, equivalently, any combination of $\beta^{(1)}$, $\beta^{(2)}$ and $\mu$. In other words, only single-node state changes are allowed during an infinitesimal interval $dt$.

The $q$s, Eqs.~\ref{eq:q_sim2com}a-\ref{eq:q_sim2com}c, now become
\begin{subequations}
\begin{align}
  q_{i,0}^{(1)} =~ &1 - \sum_{\scriptstyle {j\in\Gamma_{i,0}^{(1)}}} \beta^{(1)}P_{j\vert i,0}^{I\vert S}
  \\
  q_{i,g}^{(2)} =~ &1 - \sum_{\scriptstyle {\{j, k\}\in\Gamma_{i,g}^{(2)}}} \left[\beta^{(1)}\left(P_{jk\vert i,g}^{IS\vert S}+P_{jk\vert i,g}^{SI\vert S}+2P_{jk\vert i,g}^{II\vert S}\right)+g\beta^{(2)}P_{jk\vert i,g}^{II\vert S}\right]
\end{align}
\label{eq:q_sim2com_cont}
\end{subequations}

\noindent from which the term $1-q_{i,0}^{(1)}q_{i,0}^{(2)}q_{i,1}^{(2)}$, giving the probability that node $i$ gets infected, reads
\begin{equation}
  1 - q_{i,0}^{(1)}q_{i,0}^{(2)}q_{i,1}^{(2)} = \sum_{\scriptstyle {j\in\Gamma_{i,0}^{(1)}}} \beta^{(1)}P_{j\vert i,0}^{I\vert S} + \sum_{g=0,1}\sum_{\scriptstyle {\{j, k\}\in\Gamma_{i,g}^{(2)}}} \left[\beta^{(1)}\left(P_{jk\vert i,g}^{IS\vert S}+P_{jk\vert i,g}^{SI\vert S}+2P_{jk\vert i,g}^{II\vert S}\right)+g\beta^{(2)}P_{jk\vert i,g}^{II\vert S}\right]
  \label{eq:(1-qqq)_cont}
\end{equation}

The evolution of the probability $P_i^I$ for node~$i$ being infected, now takes the form
\begin{equation}
  \frac{d}{dt}P_i^I\left(t\right) = -\mu P_i^I\left(t\right)+P_i^S\left(t\right)\left(1-q_{i,0}^{(1)}q_{i,0}^{(2)}q_{i,1}^{(2)}\right)
  \label{eq:P_i_sim2com_cont}
\end{equation}

\noindent where $1-q_{i,0}^{(1)}q_{i,0}^{(2)}q_{i,1}^{(2)}$ is given by Eq.~\ref{eq:(1-qqq)_cont}.

\indent The state of a $(0,1)$-clique $\{i,j\}$ is governed by the following equations
\begin{subequations}
\begin{align}
  \notag \frac{d}{dt}P_{ij,0}^{II}(t) &=
  \notag P_{ij,0}^{SI}(t)\left(\beta^{(1)} + 1 - q_{i(j),0}^{(1)}q_{i,0}^{(2)}q_{i,1}^{(2)}\right)
  \\
  \notag &+ P_{ij,0}^{IS}(t)\left(\beta^{(1)} + 1 - q_{j(i),0}^{(1)}q_{j,0}^{(2)}q_{j,1}^{(2)}\right)
  \\
  &- P_{ij,0}^{II}(t)~2\mu
  \\
  \notag \frac{d}{dt}P_{ij,0}^{IS}(t) &= P_{ij,0}^{SS}(t)\left(1-q_{i(j),0}^{(1)}q_{i,0}^{(2)}q_{i,1}^{(2)}\right)
  \\
  \notag &- P_{ij,0}^{IS}(t)\left(\mu + \beta^{(1)} + 1 - q_{j(i),0}^{(1)}q_{j,0}^{(2)}q_{j,1}^{(2)}\right)
  \\
  &+ P_{ij,0}^{II}(t)~\mu
\end{align}
\label{eq:edge_sim2com_cont}
\end{subequations}

\noindent where $q_{i(j),0}^{(1)}$ coincides with $q_{i,0}^{(1)}$ except for excluding the $(0,1)$-clique $\{i,j\}$ from the sum, and analogously for the other similar terms.

The state of a $(0,2)$-clique $\{i,j,k\}$ follows the equations
\begin{subequations}
\begin{align}
  \notag \frac{d}{dt}P_{ijk,0}^{III}(t) &=
  \notag P_{ijk,0}^{SII}(t)\left(2\beta^{(1)} + 1 - q_{i,0}^{(1)}q_{i\left(jk\right),0}^{(2)}q_{i,1}^{(2)}\right)
  \\
  \notag &+ P_{ijk,0}^{ISI}(t)\left(2\beta^{(1)} + 1 - q_{j,0}^{(1)}q_{j\left(ik\right),0}^{(2)}q_{j,1}^{(2)}\right)
  \\
  \notag &+ P_{ijk,0}^{IIS}(t)\left(2\beta^{(1)} + 1 - q_{k,0}^{(1)}q_{k\left(ij\right),0}^{(2)}q_{k,1}^{(2)}\right)
  \\
  &- P_{ijk,0}^{III}(t)~3\mu
  \\
  \notag \frac{d}{dt}P_{ijk,0}^{IIS}(t) &=
  \notag P_{ijk,0}^{SIS}(t)\left(\beta^{(1)} + 1 - q_{i,0}^{(1)}q_{i\left(jk\right),0}^{(2)}q_{i,1}^{(2)}\right)
  \\
  \notag &+ P_{ijk,0}^{ISS}(t)\left(\beta^{(1)} + 1 - q_{j,0}^{(1)}q_{j\left(ik\right),0}^{(2)}q_{j,1}^{(2)}\right)
  \\
  \notag &- P_{ijk,0}^{IIS}(t)\left(2\mu + 2\beta^{(1)} + 1 - q_{k,0}^{(1)}q_{k\left(ij\right),0}^{(2)}q_{k,1}^{(2)}\right)
  \\
  &+ P_{ijk,0}^{III}(t)~\mu
  \\
  \notag \frac{d}{dt}P_{ijk,0}^{ISS}(t) &= P_{ijk,0}^{SSS}(t)\left(1-q_{i,0}^{(1)}q_{i\left(jk\right),0}^{(2)}q_{i,1}^{(2)}\right)
  \\
  \notag &- P_{ijk,0}^{ISS}(t)\left(\mu + 2\beta^{(1)} + 2 - q_{j,0}^{(1)}q_{j\left(ik\right),0}^{(2)}q_{j,1}^{(2)} - q_{k,0}^{(1)}q_{k\left(ij\right),0}^{(2)}q_{k,1}^{(2)}\right)
  \\
  \notag &+ P_{ijk,0}^{ISI}(t)~\mu
  \\
  \notag &+ P_{ijk,0}^{IIS}(t)~\mu
\end{align}
\label{eq:triangle0_sim2com_cont}
\end{subequations}

\noindent where $q_{i(jk),0}^{(2)}$ coincides with $q_{i,0}^{(2)}$ except for excluding the $(0,2)$-clique $\{i,j,k\}$ from the sum, and analogously for the other similar terms.

Finally, for a $(1,2)$-clique $\{i,j,k\}$, we get the following equations
\begin{subequations}
\begin{align}
  \notag \frac{d}{dt}P_{ijk,1}^{III}(t) &=
  \notag P_{ijk,1}^{SII}(t)\left(2\beta^{(1)} + \beta^{(2)} + 1 - q_{i,0}^{(1)}q_{i,0}^{(2)}q_{i\left(jk\right),1}^{(2)}\right)
  \\
  \notag &+ P_{ijk,1}^{ISI}(t)\left(2\beta^{(1)} + \beta^{(2)} + 1 - q_{j,0}^{(1)}q_{j,0}^{(2)}q_{j\left(ik\right),1}^{(2)}\right)
  \\
  \notag &+ P_{ijk,1}^{IIS}(t)\left(2\beta^{(1)} + \beta^{(2)} + 1 - q_{k,0}^{(1)}q_{k,0}^{(2)}q_{k\left(ij\right),1}^{(2)}\right)
  \\
  &- P_{ijk,1}^{III}(t)~3\mu
  \\
  \notag \frac{d}{dt}P_{ijk,1}^{IIS}(t) &=
  \notag P_{ijk,1}^{SIS}(t)\left(\beta^{(1)} + 1 - q_{i,0}^{(1)}q_{i,0}^{(2)}q_{i\left(jk\right),1}^{(2)}\right)
  \\
  \notag &+ P_{ijk,1}^{ISS}(t)\left(\beta^{(1)} + 1 - q_{j,0}^{(1)}q_{j,0}^{(2)}q_{j\left(ik\right),1}^{(2)}\right)
  \\
  \notag &- P_{ijk,1}^{IIS}(t)\left(2\mu + 2\beta^{(1)} + \beta^{(2)} + 1 - q_{k,0}^{(1)}q_{k,0}^{(2)}q_{k\left(ij\right),1}^{(2)}\right)
  \\
  &+ P_{ijk,1}^{III}(t)~\mu
  \\
  \notag \frac{d}{dt}P_{ijk,1}^{ISS}(t) &= P_{ijk,1}^{SSS}(t)\left(1-q_{i,0}^{(1)}q_{i,0}^{(2)}q_{i\left(jk\right),1}^{(2)}\right)
  \\
  \notag &- P_{ijk,1}^{ISS}(t)\left(\mu + 2\beta^{(1)} + 2 - q_{j,0}^{(1)}q_{j,0}^{(2)}q_{j\left(ik\right),1}^{(2)} - q_{k,0}^{(1)}q_{k,0}^{(2)}q_{k\left(ij\right),1}^{(2)}\right)
  \\
  \notag &+ P_{ijk,1}^{ISI}(t)~\mu
  \\
  \notag &+ P_{ijk,1}^{IIS}(t)~\mu
\end{align}
\label{eq:triangle1_sim2com_cont}
\end{subequations}

\noindent where $q_{i(jk),1}^{(2)}$ coincides with $q_{i,1}^{(2)}$ except for excluding the $(1,2)$-clique $\{i,j,k\}$ from the sum, and analogously for the other similar terms.

Evidently, second-order dynamical correlations let $\beta^{(2)}$ appear in the dynamical equations. When linearizing Eqs.~\ref{eq:q_sim2com_cont}-\ref{eq:triangle1_sim2com_cont} as done in Methods, the products between a state probability with some infected node and $1-q_{i,0}^{(1)}q_{i,0}^{(2)}q_{i,1}^{(2)}$ (et similia) disappear as negligible terms, while those ones corresponding to infections within the considered triangles do not. Consequently, $\beta^{(2)}$ does show up in the final eigenvalue equation providing the critical point, $\beta^{(1)}_{\textup{cr}}$. In Supplementary Fig.~\ref{fig:beta_cr_DM_cont} we report the dependence of $\beta^{(1)}_{\textup{cr}}$ on $\beta^{(2)}$ as computed for a Dorogovtsev-Mendes SC.

\vspace{2.ex}
\noindent{\fontfamily{cmss}\selectfont{\textbf{Non-0-connected structures.}}}
Eq.~\ref{eq:(1-qqq)_cont} (or Eq.~\ref{eq:q_sim2com_cont}b) holds for a $0$-connected SC. Nonetheless, differently from its discrete-time version, it can be adapted to hold for any connectedness. Indeed, the sum over the $(\cdot,3)$-cliques in Eq.~\ref{eq:(1-qqq)_cont} (or Eq.~\ref{eq:q_sim2com_cont}b) can now be split into two sums: one regarding the first-order infections coming from the edges of the $(\cdot,3)$-cliques, and one regarding the second-order infections coming from the $(1,3)$-cliques. In the former, rather than over the couples of neighbors of a node, one can sum over its single neighbors, in this way avoiding to over-count their contribution. To this end, one may define $\tilde{\Gamma}_{i}^{(2)}$ as the set of neighbors of node $i$ such that $j\in\tilde{\Gamma}_{i}^{(2)}$ if $\exists k$ such that $\{j,k\}\in\Gamma_{i,g}^{(2)}$, $\forall g\in\{0,1\}$. Relaxing the edge-disjoint requirement, we can now look for a standard edge clique cover of the structure. Keeping the $(g,n)$ nomenclature for the cliques in the cover, Eq.~\ref{eq:(1-qqq)_cont} adopts the form
\begin{equation}
  1 - q_{i,0}^{(1)}q_{i,0}^{(2)}q_{i,1}^{(2)} = \sum_{\scriptstyle {j\in\Gamma_{i,0}^{(1)}}} \beta^{(1)}P_{j\vert i,0}^{I\vert S} + \sum_{\scriptstyle {j\in\tilde{\Gamma}_{i}^{(2)}}} \beta^{(1)}\tilde{P}_{j\vert i}^{I\vert S} + \sum_{\scriptstyle {\{j, k\}\in\Gamma_{i,1}^{(2)}}} \beta^{(2)}P_{jk\vert i,g}^{II\vert S}
  \label{eq:(1-qqq)_cont_general}
\end{equation}
\noindent where $\tilde{P}_{j\vert i}^{I\vert S}\equiv P_{jk\vert i,g}^{II\vert S}+P_{jk\vert i,g}^{IS\vert S}$ for $k$ such that $\{j,k\}\in\Gamma_{i,g}^{(2)}$. Since the edge $\{i,j\}$ is now allowed to be included in more than one $(\cdot,3)$-clique, $k$ can identify more than one node. Differently from single nodes, a complete marginalization of the dynamical equation for the edge $\{i,j\}$, that would make $\tilde{P}_{j\vert i}^{I\vert S}$ independent from the chosen $k$, is however unfeasible (and not only with the chosen closure, Eq.~(7)). Therefore, a more symmetrical way to evaluate $\tilde{P}_{j\vert i}^{I\vert S}$ is to compute it as the average value of $P_{jk\vert i,g}^{II\vert S}+P_{jk\vert i,g}^{IS\vert S}$ over all nodes $k$ such that $\{j,k\}\in\Gamma_{i,g}^{(2)}$, $g=0,1$.

What has been done here for simplicial $2$-complexes, can be extended to define the continuous-time limit of the MECLE for non-$0$-connected simplicial complexes of any higher dimension.

\newpage

\stepcounter{SMfigure}
\begin{figure}[ht!]
  \centering
  \includegraphics[width=.85\linewidth]{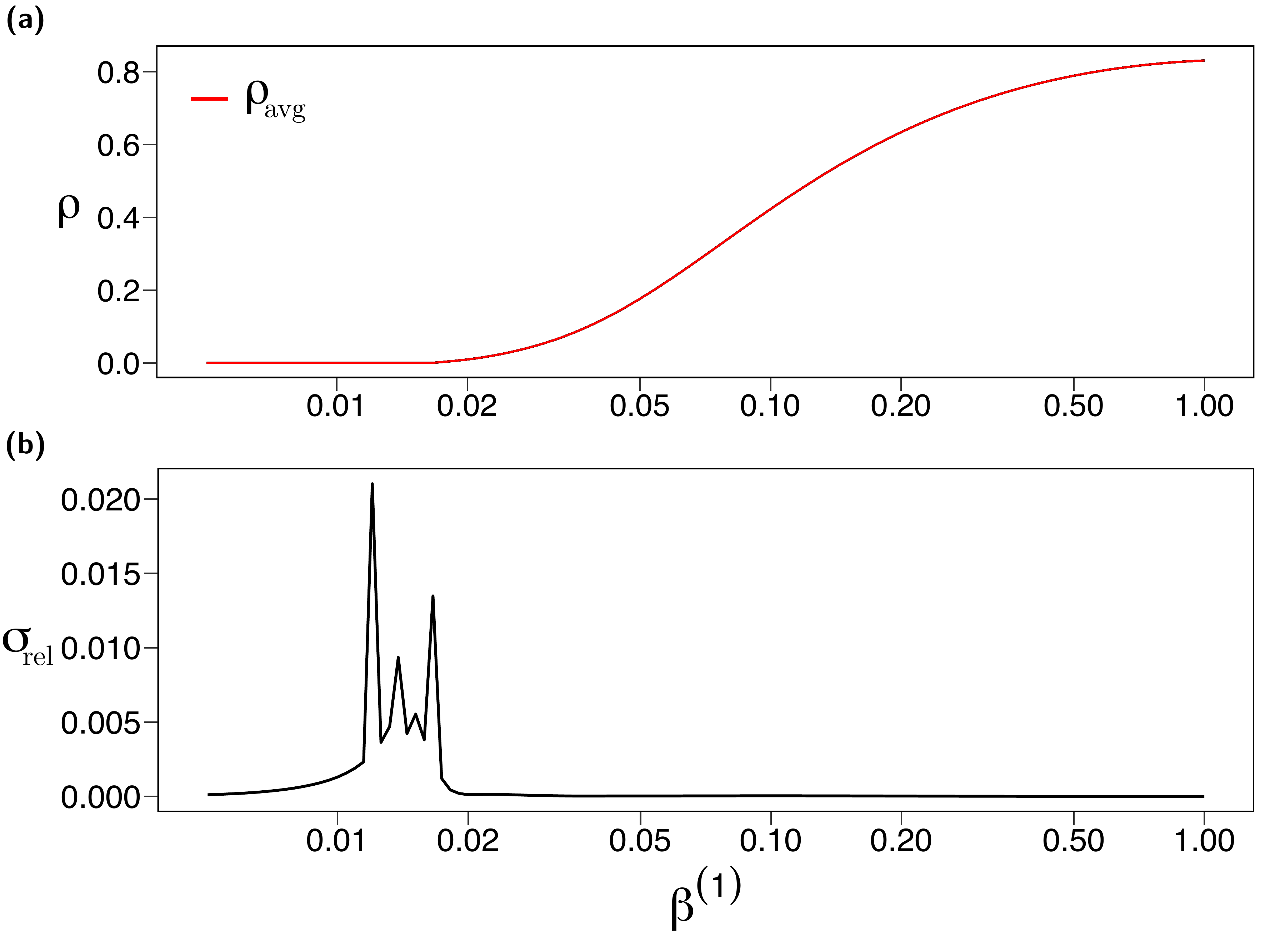}

  \caption{The robustness of the MECLE under different EECCs is assessed for a graph generated from the Dorogovtsev-Mendes model, forming a 0-connected simplicial complex. This boasts a very high rate of edge overlap and, therefore, potentially different EECCs could lead the MECLE to make notably dissimilar predictions. Indeed, about $34\%$ of the edges in the network are shared at least by two maximal cliques. Specifically, within that $34\%$, edges are shared by about $2.5$ maximal cliques on average, with a standard deviation of $0.85$ and a skewness of $2.3$. The recovery probability is $\mu=0.2$. (a) $20$~prevalence curves obtained by $20$~different EECCs are shown in black, while in red is reported their average curve $\rho_\textup{avg}(\beta^{(1)})$, defined by the average value of $\rho$ at each value of $\beta^{(1)}$. Since the deviations are below the width of the red line, the black lines are not visible. (b) The ratio $\sigma_\textup{rel}$ between the sample standard deviation $\sigma$ and $\rho_\textup{avg}$ for each $\beta^{(1)}$. The value of $\sigma_\textup{rel}$ vanishes everywhere, expect for a small region around the epidemic threshold. This is below $2.5\%$ for values of $\rho_\textup{avg}\sim 10^{-4}\sim N^{-1}$, proving the remarkable robustness of the model.}
  \label{fig:robustness_0con}
\end{figure}

\stepcounter{SMfigure}
\begin{figure}[ht!]
  \centering
  \includegraphics[width=.85\linewidth]{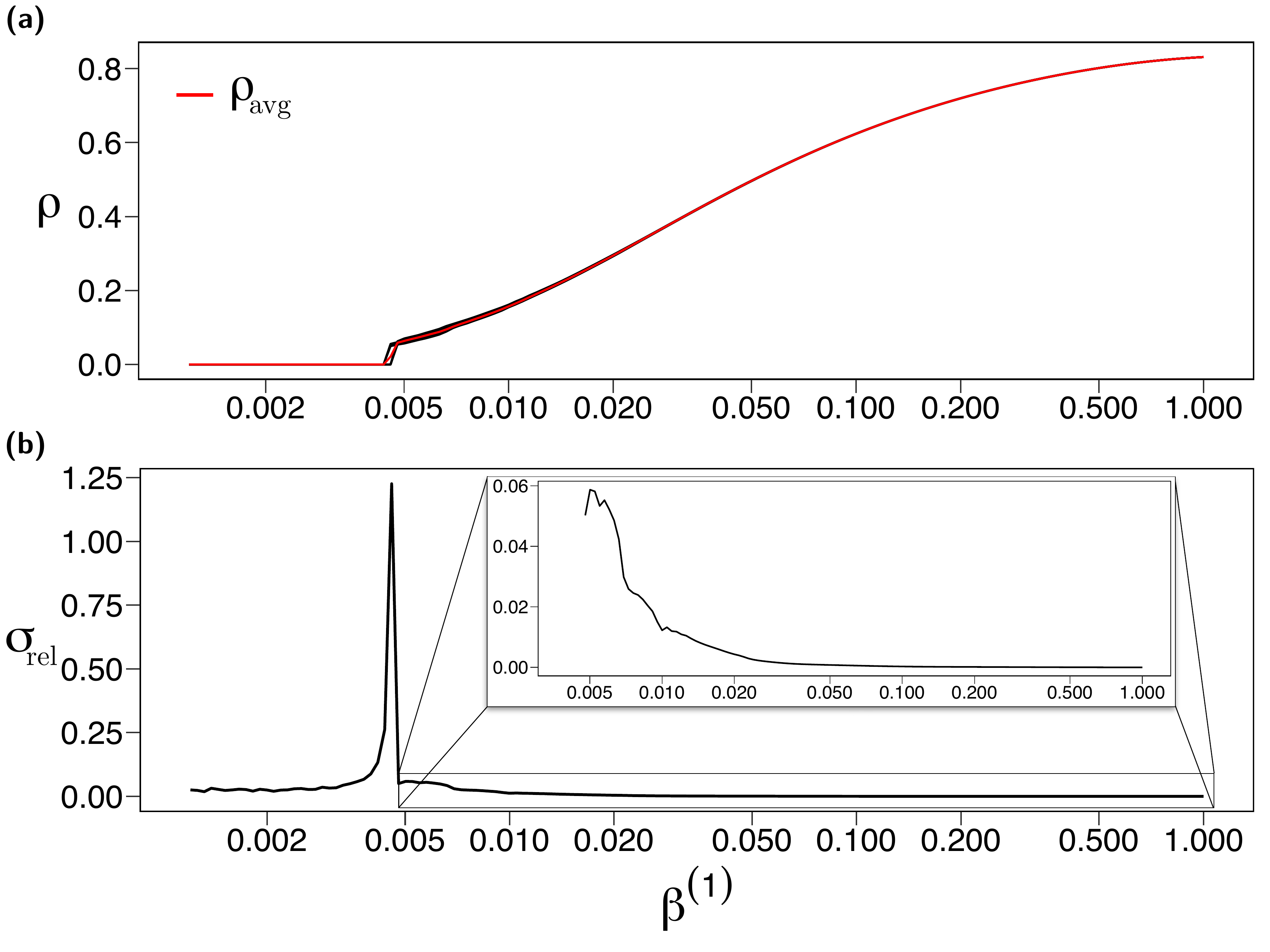}

  \caption{The robustness of the MECLE under different EECCs is assessed for a non-0-connected simplicial complex generated from the Dorogovtsev-Mendes network used in Supplementary Fig.~\ref{fig:robustness_0con}. The recovery probability is $\mu=0.2$. (a) $20$ prevalence curves obtained by $20$ different EECCs are shown in black, while in red is reported their average curve $\rho_\textup{avg}(\beta^{(1)})$, defined by the average value of $\rho$ at each value of $\beta^{(1)}$. (b) The ratio $\sigma_\textup{rel}$ between the sample standard deviation $\sigma$ and $\rho_\textup{avg}$ for each $\beta^{(1)}$. The pick at about $\beta^{(1)}=0.0047$ is due to some curves transitioning at slightly different values of $\beta^{(1)}$. In fact, the uncertainty about the location of the critical point is of only about $0.0002$, corresponding to a relative uncertainty of less than $5\%$. The inset plot shows a zoom of $\sigma_\textup{rel}$ to the right of the transition:  $\sigma_\textup{rel}$ stays below the $6\%$ next to the transition, while rapidly decreasing towards zero for larger $\beta^{(1)}$s, hence enlightening the robustness of the model.}
  \label{fig:robustness_non0con}
\end{figure}

\stepcounter{SMfigure}
\begin{figure}[ht!]
  \centering
  \includegraphics[width=.90\linewidth]{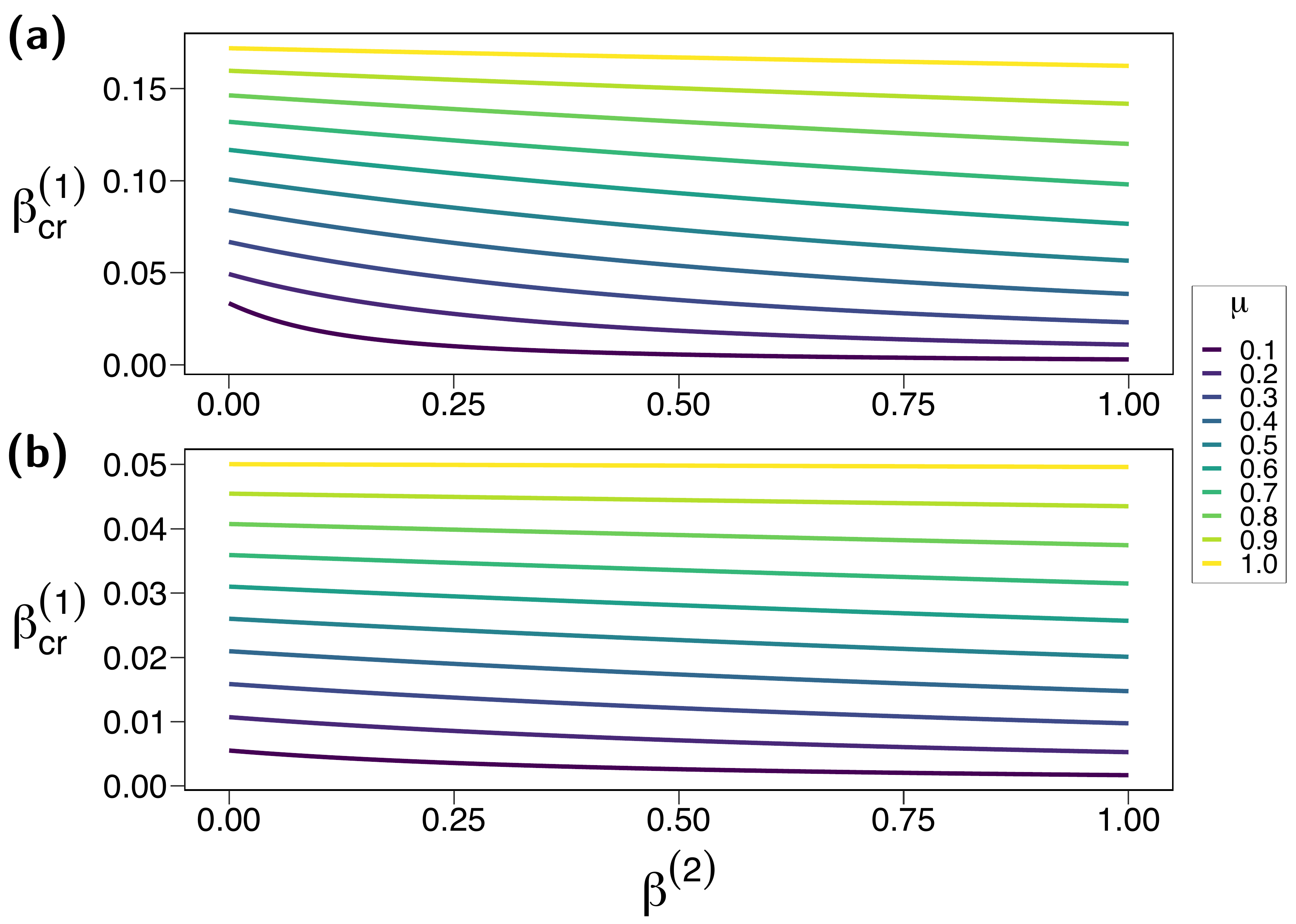}

  \caption{The value of the epidemic threshold $\beta^{(1)}_{\textup{cr}}$, as computed from Eq.~(\ref{eq:epi_th_regular}), is shown against $\beta^{(2)}$ for (a) any regular clique $2$-complex with $k^{(0,1)}=k^{(0)}=0$ and $k^{(1,2)}=k^{(1)}=3$ (the periodic triangular clique complex in the  main text falls within this class); (b) any regular clique $2$-complex with $k^{(0,1)}=k^{(0)}=12$ and $k^{(1,2)}=k^{(1)}=4$ (a proxy for large random SCs). Note that the mean-field approximation, Eq.~(22), wrongly predicts $\beta^{(1)}_{\textup{cr}}=\mu/k$, $\forall\beta^{(2)}$, where $k=6$ in (a), $k=20$ in (b).}
  \label{fig:regular_diagram}
\end{figure}

\stepcounter{SMfigure}
\begin{figure}[ht!]
  \centering
  \includegraphics[width=.90\linewidth]{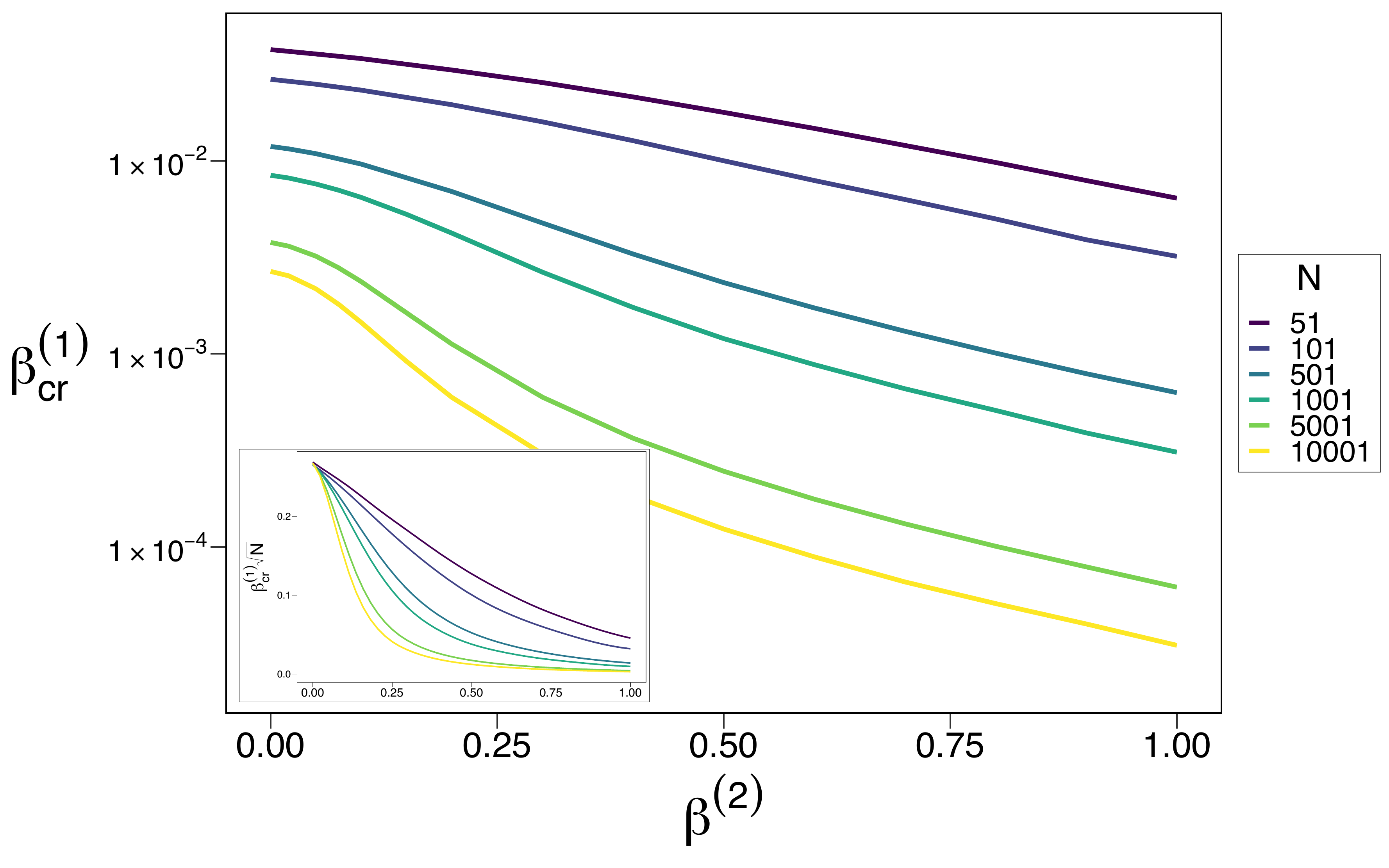}

  \caption{The value of the epidemic threshold $\beta^{(1)}_{\textup{cr}}$ is shown against $\beta^{(2)}$ for simplicial $2$-complexes constructed from the Friendship graph $\textup{F}_n$, a proxy for extremely heterogeneous structures, where $n=\frac{N-1}{2}$. Here $p_\triangle=0.5$, thus $k^{(0,2)}=k^{(1,2)}=n/2$ for the central node, $k^{(1,2)}=1$ and $k^{(0,2)}=0$ for half of the peripheral nodes, and $k^{(0,2)}=0$ and $k^{(1,2)}=1$ for the other half. The recovery probability is $\mu=0.2$. The mean-field approximation, Eq.~(22), wrongly predicts $\beta^{(1)}_{\textup{cr}}=\mu/{\bar k}$, $\forall\beta^{(2)}$, where $\bar k=3\frac{N-1}{N}$. The inset plot shows the curves $\beta^{(1)}_{\textup{cr}}\sqrt{N}$ vs. $\beta^{(2)}$. These all collapse to the same value $x^\ast\approx 0.27$ at $\beta^{(2)}=0$, while being smaller than $x^\ast$ for any $\beta^{(2)}>0$, hence proving that $\beta^{(1)}_{\textup{cr}}\rightarrow 0$ for $N\rightarrow \infty$, $\forall \beta^{(2)}$.}
  \label{fig:windmill_diagram}
\end{figure}

\stepcounter{SMfigure}
\begin{figure}[ht!]
  \centering
  \includegraphics[width=.85\linewidth]{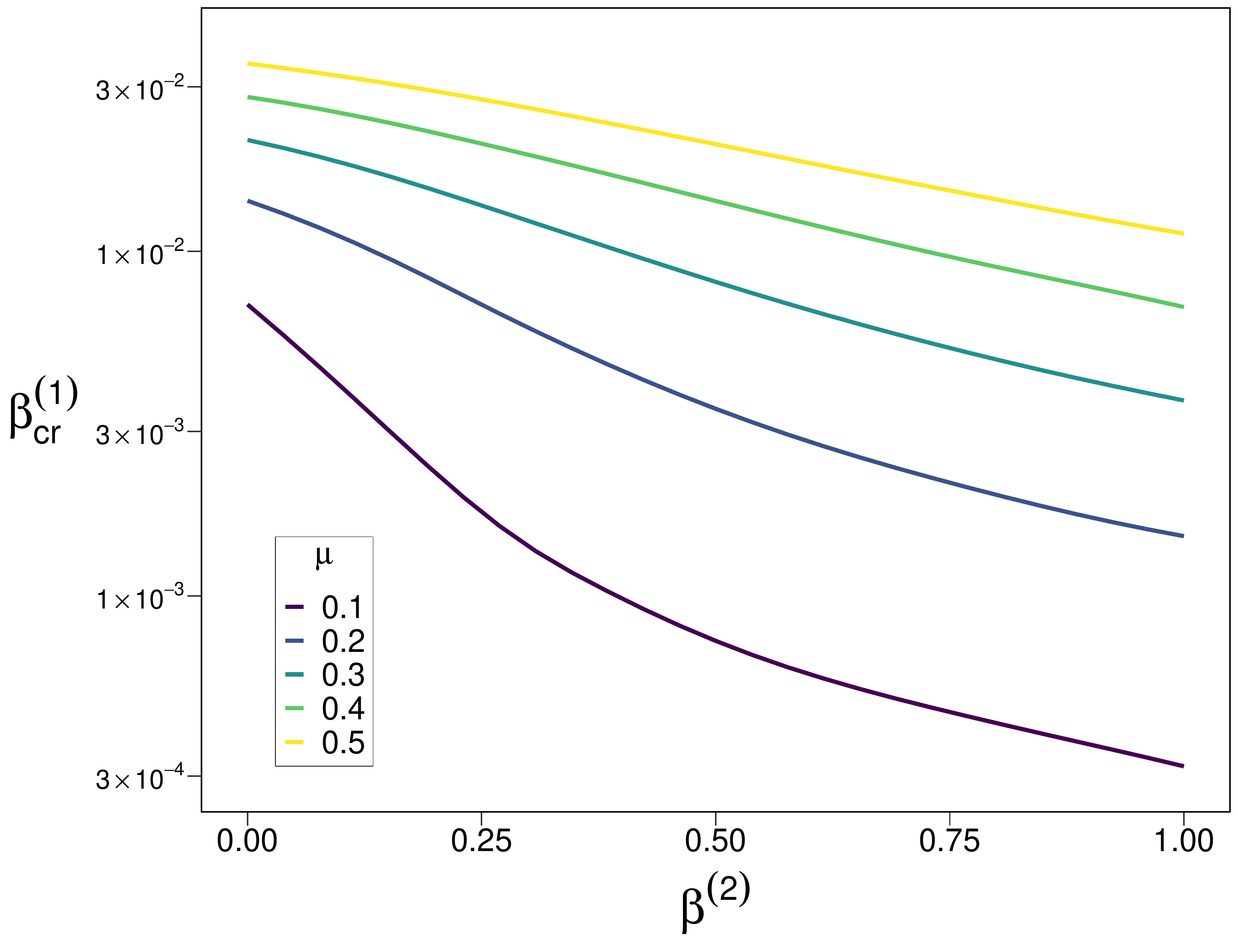}

  \caption{\rev{The value of the epidemic threshold $\beta^{(1)}_{\textup{cr}}$, as predicted in the continuous-time limit by the MECLE, is shown against $\beta^{(2)}$ for a Dorogovtsev-Mendes SC with $\bar{k}^{(0,1)}=1.10$ and $\bar{k}^{(1,2)}=1.45$. The mean-field approximation, Eq.~(22), wrongly predicts $\beta^{(1)}_{\textup{cr}}=\mu/{\bar k}=0.05$, $\forall\beta^{(2)}$.}}
  \label{fig:beta_cr_DM_cont}
\end{figure}

\end{document}